\documentclass{biophys}
\usepackage{helvet,times}
\usepackage{bm,textcomp}
\usepackage{booktabs}
\usepackage{longtable}

\jno{kxl014} 
\gridframe{N} 
\cropmark{N} 

\doi{doi: 10.1529/biophysj.XYZ.123456}

\Month{June}
\Year{2015}

\usepackage{amsmath}
\usepackage{amssymb}
\usepackage{graphicx}
\usepackage[round,numbers,sort&compress]{natbib}

\newcommand{\revision}[1]{{#1}}
\newcommand{\Na}{\text{Na}}
\newcommand{\Rb}{\text{Rb}}
\newcommand{\Sr}{\text{Sr}}
\newcommand{\Cl}{\text{Cl}}
\newcommand{\CoHex}{\text{CoHex}}
\newcommand{\ideal}{\text{id}}
\newcommand{\excess}{\text{ex}}
\newcommand{\bulk}{\text{bulk}}
\renewcommand{\vec}[1]{{\bf{#1}}}
\newcommand{\FE}{\mathcal{F}}

\begin{document}

\setcounter{page}{1} 

\title{The role of correlation and solvation in ion interactions with B-DNA}


\author{Maria~L.~Sushko\thanks{Physical Sciences Division, Pacific Northwest National Laboratory, Richland, WA 99352.},
Dennis~G.~Thomas\thanks{Biological Sciences Division, Pacific Northwest National Laboratory, Richland, WA 99352},
Suzette~A.~Pabit\thanks{School of Applied and Engineering Physics, Cornell University, Ithaca, NY 14853-3501.},
Lois~Pollack\thanks{School of Applied and Engineering Physics, Cornell University, Ithaca, NY 14853-3501.},
Alexey~V.~Onufriev\thanks{Department of Computer Science and Department of Physics, Virginia Tech, Blacksburg, VA 24061.},
Nathan~A.~Baker\thanks{Computational and Statistical Analytics Division, Pacific Northwest National Laboratory, Richland, WA 99352; Division of Applied Mathematics, Brown University.}}


\begin{abstract}%
	{The ionic atmospheres around nucleic acids play important roles in biological function.
	Large-scale explicit solvent simulations coupled to experimental assays such as anomalous small-angle X-ray scattering (ASAXS) can provide important insights into the structure and energetics of such atmospheres but are time- and resource-intensive.
	In this paper, we use classical density functional theory (cDFT) to explore the balance between ion-DNA, ion-water, and ion-ion interactions in ionic atmospheres of RbCl, SrCl$_2$, and CoHexCl$_3$ (cobalt hexammine chloride) around a B-form DNA molecule.
	The accuracy of the cDFT calculations was assessed by comparison between simulated and experimental ASAXS curves, demonstrating that an accurate model should take into account ion-ion correlation and ion hydration forces, DNA topology, and the discrete distribution of charges on  {the DNA backbone} . As expected, these calculations revealed significant differences between monovalent, divalent, and trivalent cation distributions around DNA.
	About half of the DNA-bound Rb$^+$ ions penetrate into the minor groove of the DNA and half adsorb on the DNA {backbone} .
	The fraction of cations in the minor groove decreases for the larger Sr$^{2+}$ ions and becomes zero for CoHex$^{3+}$ ions, which all adsorb on the DNA {backbone} .
	The distribution of CoHex$^{3+}$ ions is mainly determined by Coulomb and steric interactions, while ion-correlation forces play a central role in the monovalent Rb$^+$ distribution and a combination of ion-correlation and hydration forces affect the Sr$^{2+}$ distribution around DNA.
	{This does not imply that correlations in CoHex solutions are weaker or stronger than for other ions.
	Steric inaccessibility of the grooves to large CoHex ions leads to their binding at the DNA surface.
	In this binding mode, first-order electrostatic interactions (Coulomb) dominate the overall binding energy as evidenced by low sensitivity of ionic distribution to the presence or absence of second-order electrostatic correlation interactions.}}
	{}
	{Please address correspondence to Nathan Baker (nathan.baker@pnnl.gov).}
\end{abstract}

\maketitle 

\section*{Introduction}

Interactions with ions stabilize nucleic acid secondary and tertiary structure, have a major impact on DNA packing in cells, and strongly influence protein and drug binding \cite{Rau:1984, Rau:1984a, Knobler:2009, Xiang:2009, Draper:2008, Draper:2005, Woodson:2005, Misra:1994, Misra:1994a}.
A fraction of counterions bind to specific sites on nucleic acids and can be detected in crystallographic structures \cite{Frederiksen:2009}, while other counterions form a dynamic ion atmosphere around DNA, diffusing along the molecule and exchanging with ions in bulk solution \cite{Freisinger:2007}.
Mean-field approaches such as Manning counterion condensation \cite{Manning:1978} and Poisson-Boltzmann (PB) \cite{Lamm:2003, Baker:2004, Fixman:1979, Anderson:1990} theory have been used to obtain insight into ion distributions around biomolecules and ion-mediated interactions between macro-ions {and have been compared with experimental data with some success \cite{Chu:2007, Giambasu:2014, Pabit:2009, Bai:2007}.}
While successful in describing some properties of nucleic acids in electrolyte solutions (e.g., RNA p$K_a$ shifts \cite{Tang:2007}, monovalent ion concentration linkages to ligand-DNA binding \cite{Shkel:2012, Misra:1994, Misra:1994a, Misra:1998}, and low valency ion distributions around DNA), these mean-field methods often fail when the ion charge concentration increases.
For example, PB models cannot capture the displacement of Na$^+$ by Mg$^{2+}$ around DNA in mixed solutions \cite{Chu:2007} or ion-mediated DNA-DNA attractive interactions \cite{Kornyshev:1999}.
By imposing the constraint that a fraction of the counterions are bound (condensed) to polyelectrolyte and part form ionic atmosphere in the mean-field counterion condensation theory,
it has been possible to reproduce attraction between like-charged polyelectrolytes in the presence of monovalent counterions in the intermediate range of separations \cite{Kornyshev:1999, Ray:1994, Manning:2011, Perico:2011, Pietronave:2008}.
Manning suggests that the origin of this effect lies in the increase in entropy due to the increase in the effective volume available for condensed counterions as two DNA molecules approach \cite{Manning:2011}.
Such condensation implies penetration of ions through the DNA hydration layer and their partial desolvation to form direct bonds with DNA \cite{Long:2006, Allahyarov:2003, Giambasu:2014}.
Describing this process requires atomistic or coarse-grained representation of the macro-ion, which captures both the discreteness of charge distribution on the DNA {backbone}  and DNA topology, as well as a model for ion desolvation.
Such characteristics are not currently present in the PB equation or other popular models of biomolecular electrostatics.

These failures suggest that, to reliably describe ion distribution around nucleic acids, the theoretical model must be refined to include more detailed interactions and incorporate higher-order non-mean-field interactions such as fluctuations.
Such extensions of PB approach have been developed for simple geometries (e.g., plates, rods, spheres, etc.) to include second-order terms representing the interactions between fluctuations in ionic densities \cite{GonzalezAmezcua:2001, Kjellander:1988, Blum:1975, Henderson:1978, Hoye:1978, Jiang:2001, Jiang:2001a}.
These extended models and molecular simulations \cite{Guldbrand:1986, Rouzina:1996, Rouzina:1996a, Arenzon:1999, Levin:1999, Shklovskii:1999, Shklovskii:1999a, Netz:2001, Naji:2004} as well as experimental data \cite{Danilowicz:2009, Lin:1978, Drifford:1985, Zero:1984, Matsuoka:1991, Sedlak:1993, Sedlak:1994, Sedlak:1996, Sedlak:1997, Sedlak:1992, Sedlak:1992a, Bockstaller:2001} predict attraction between like-charged objects in the presence of multivalent electrolytes.

In this study, we establish a minimal model based on classical density functional theory (cDFT) to systematically study the influence of the discrete DNA molecular charge representation, ion-ion correlations, and ion-solvent interactions on the distribution of monovalent and multivalent ions around highly charged macromolecules.
We show that this model is able to accurately reproduce the results of anomalous small-angle X-ray scattering (ASAXS) experiments \cite{Pabit:2010, Andresen:2008, Pabit:2009, Pabit:2009a} for B-DNA in RbCl, SrCl$_2$ and CoHexCl$_3$ solutions.
As expected, ion-ion correlations play a significant role in the accurate prediction of ASAXS curves.
However, our results also demonstrate the importance of ion solvation in cation-DNA interactions and show that for doubly-charged cations these interactions can be as important as ion-ion correlations in modeling ion distributions around DNA.

\section*{Methods}

\subsection*{DNA models}
\begin{figure}[t]
	\centering
	(a)~\includegraphics[keepaspectratio,width=0.6in]{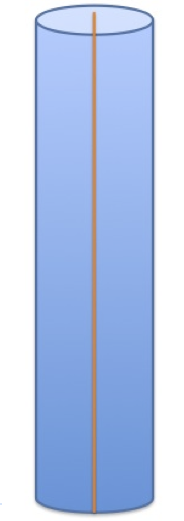}
	(b)~\includegraphics[keepaspectratio,width=1in]{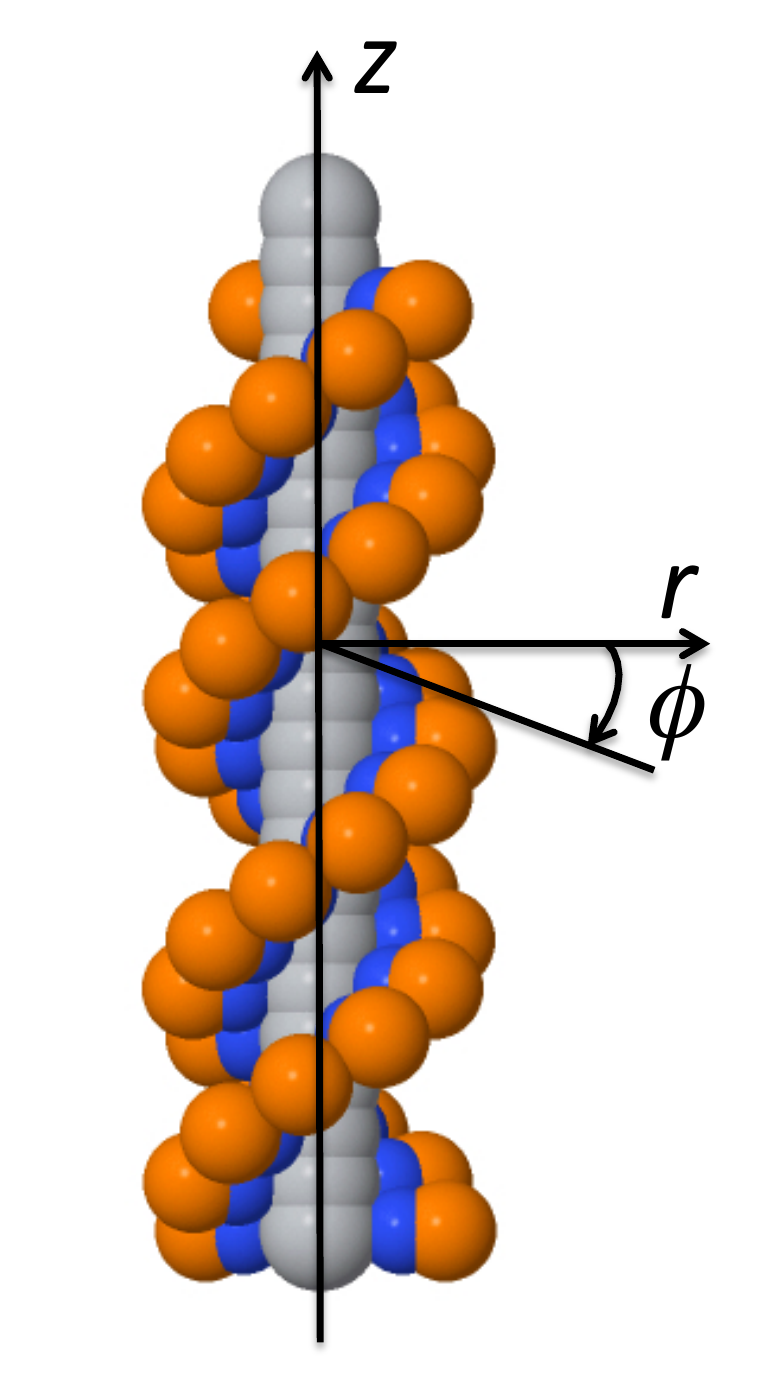}
	\caption{Macroion models used in classical DFT simulations: (a) Model of a cylinder with uniform axial charge density; (b) discrete charge model.}
	\label{fig:dna-models}
\end{figure}

We used two coarse-grained models for the DNA macro-ion in the cDFT simulations: an infinitely long cylinder with a uniform line charge density along its $z$-axis (charge distribution -1 e per 0.17 nm and the 2 nm cylinder diameter) and a model with a discrete charge distribution (Fig.~\ref{fig:dna-models}).
{Na$^+$ counterions present at 0.78 M concentration in all DNA calculations.}
The discrete charge distribution of the second model is described by three particle types:  two helical arrays of charged spheres that represent the phosphate groups (charge -1 e, diameter 0.42 nm), two helical arrays of neutral spheres (diameter 0.42 nm) that represent the sugar/base groups, and an array of overlapping neutral spheres (diameter 0.78 nm) defining the DNA axis \cite{Allahyarov:2005}.
The positions of these spheres were chosen to mimic B-form DNA using a cylindrical coordinate system $\left( r^s_j, \phi^s_j, z^s_j \right)$ for DNA {backbone} $s$ and {base pair} $j$.
The phosphate spheres have coordinates $r^s_j = 0.89$ nm, $\phi_j^s = \phi^s_0 + 36 j$ degrees, and $z_j^s = z_0^s + 0.34 j$ nm; the sugar/base spheres have coordinates $r_i^s = 0.59$ nm, $\phi_j^s = \phi^s_0 + 36 j$ degrees, and $z_j^s = z_0^s + 0.34 j$ nm; and the axis spheres have coordinates $r = 0$ nm, $\phi = 0$ degrees, and $z_j = 0.5 + 0.34 j$ nm.
There are {10 base pairs} $(j = 0, \ldots, 9)$ per turn of B-DNA; the angular cylindrical coordinates for {backbone} start at $\phi^{(1)}_0 = 0$ and $\phi^{(2)}_0 = 154$ degrees, respectively.

\subsection*{Computational models}
A variety of computational models were used with the DNA models described above to assess the influence of different energetic contributions on DNA-ion interactions.
These models are summarized in Table \ref{tab:models} and described in detail in the following sections.
\begin{table}
	\caption{A summary of the different computational models used in this paper to assess the influence of different energetic contributions (ion-ion electrostatic correlations, ion-ion steric correlations, ion-solvent interactions, and water structural changes) on DNA-ion distributions and compare the resulting distribution functions with experimental ASAXS data.
	The rows provide model descriptions while the columns indicate which physical phenomena are included by the models.
	{Additional quantitative data on model results are included in Table \ref{tab:results}.}}
	\label{tab:models}
	\centering
	\begin{tabular}{p{1.0in}|p{0.6in}p{0.6in}p{0.6in}p{0.6in}|p{0.6in}p{0.6in}p{0.6in}}
		\toprule
		Model & Ion-ion elec.\ correl. & Ion-ion steric correl. & Ion-solvent interact. & Water struct. change & Rb$^+$ exp.\ agreement & Sr$^{2+}$ exp.\  agreement & CoHex$^{3+}$ exp.\ agreement \\
		\hline
		NLPB & no & no & no & no & no* & no* & yes \\
		cDFT, no correlation & no & yes & no & yes & no* & no* & yes \\
		cDFT, no ion solvation & yes & yes & no & yes & yes & no* & yes \\
		cDFT, full model & yes & yes & yes & yes & yes & yes & yes \\
		\bottomrule
	\end{tabular} \\
	*Agreement with experiment can be obtained by fitting ion radii.
\end{table}

\subsubsection*{Classical density functional theory (cDFT)}
Classical DFT (cDFT) has been previously used to determine the equilibrium distributions of multicomponent salt solutions surrounding DNA \cite{Wu:2007, Cao:2005, Ovanesyan2014}.
In our cDFT models, the aqueous salt solution was modeled as a dielectric medium with $\epsilon = 78.5$, charged spherical particles representing ions, and neutral spherical particles representing water molecules.
The concentration of spherical ``water molecules" was 55.5 M, chosen to model experimental water density.
The solutions considered in this work were aqueous NaCl, RbCl, SrCl$_2$ and CoHexCl$_3$ electrolytes in Na$^+$ buffer.
These electrolytes were chosen based on the availability of the experimental data for these systems \cite{Pabit:2010, Andresen:2008}.
We used experimental crystalline ionic diameters for mobile ions: $\sigma_{\Na} = 0.204$ nm, $\sigma_{\CoHex} = 1.166$ nm, $\sigma_{\Sr} = 0.252$ nm, $\sigma_{\Rb}$ = 0.322 nm, $\sigma_{\text{Cl}} = 0.362$ nm, and $\sigma_{\text{water}} = 0.275$ nm \cite{Marcus:1988}.
The ion charges were $q_{\Na} = +1$, $q_{\CoHex} = +3$, $q_{\Sr} = +2$, $q_{\Rb} = +1$, $q_{\text{Cl}}$ = -1, and $q_{\text{water}} = 0$.
Parameterization of the cDFT model was performed against experimental data for the concentration dependence of mean activity coefficients in bulk electrolyte solutions (see Supporting Information).
All calculations were performed at 298 K temperature.

To determine the equilibrium water and ion distributions via cDFT, the total Helmholtz free energy functional is minimized with respect to the densities of all the species in the presence of rigid DNA models.
For this optimization, it is convenient to partition the total free energy of the system into so-called ideal ($\FE^{\text{id}}$) and excess components ($\FE^{\text{ex}}$) \cite{Wu:2007}.
The ideal free energy corresponds to the non-interacting system and is determined by the configurational entropy contributions from water and small ions,
\begin{equation}
	\FE^{\ideal} =
	k T \sum_i^N \int_\Omega \left( \rho_i(\vec{r}) \log \rho_i(\vec{r}) - \rho_i(\vec{r}) \right) d \vec{r}
\end{equation}
where $k$ is Boltzmann's constant, $T$ is the temperature, $\rho_i : \Omega \mapsto [0,1]$ is the density profile of ion species $i$, $N$ is the number of ion species, $\vec{r} \in \Omega$ is the ion coordinate, and $\Omega \in \mathbb{R}^3$ is the calculation domain.
The excess free energy is generally not known exactly but can be approximated by
\begin{equation}
	\FE^{\excess} \approx \FE^{\excess}_{\text{hs}} + \FE^{\excess}_{\text{corr}} + \FE^{\excess}_{\text{C}} + \FE^{\excess}_{\text{solv}}
\end{equation}
where $\FE^{\excess}_{\text{hs}}$ is the hard-sphere repulsion term, $\FE^{\excess}_{\text{corr}}$ is the {ion-ion electrostatic correlation} term, $\FE^{\excess}_{\text{C}}$ is the direct Coulomb term, and $\FE^{\excess}_{\text{solv}}$ is the {ion-solvent interaction} term included in some cDFT calculations (as described below).

The {ion-ion steric correlation} term excess free energy describes ion and water many-body interactions in condensed phase due to density fluctuations and can be approximated by Fundamental Measure Theory \cite{Yu:2002} as
\begin{equation}
	\FE^{\excess}_{\text{hs}} \approx \int_\Omega \Phi^{\text{hs}}\left[ n_w \left( \vec{r} \right) \right] d \vec{r}
\end{equation}
where the functional $\Phi^{\text{hs}}$ has the form \cite{Da:2014}
\begin{multline}
	\Phi_{\text{hs}}(\boldsymbol{r}) =  -n_0\ln(1-n_3)+\frac{n_1n_2}{1-n_3} +
	\left[\frac{1}{36\pi n_3^2}\ln(1-n_3) + \frac{1}{36\pi n_3(1-n_3)^2} \right] n_2^3 \\
	- \frac{\boldsymbol{n}_1\cdot\boldsymbol{n}_2}{1-n_3}
	- \left[ \frac{1}{12\pi n_3^2}\ln(1-n_3)+\frac{1}{12\pi n_3(1-n_3)^2} \right] n_2(\boldsymbol{n}_2\cdot\boldsymbol{n}_2).
	\label{hard-sphere-weighted-density-modified}
\end{multline}
where $n_\alpha$ and $\boldsymbol{n}_\beta$ are the scalar and vector weighted averages of the density distribution functions $\rho_i(\boldsymbol{r})$ and are defined by:
\begin{align*}
	n_\alpha (\boldsymbol{r}) & = \sum_i \int_\Omega \rho_i(\boldsymbol{r}')\omega_i^{(\alpha)}(\boldsymbol{r}'-\boldsymbol{r})\,{\rm d}\boldsymbol{r}',\;\; \text{~for~} \alpha = 0, 1, 2, 3 \\
	\boldsymbol{n}_\beta(\boldsymbol{r}) & = \sum_i \int_\Omega \rho_i(\boldsymbol{r}')\boldsymbol{\omega}_i^{(\beta)}(\boldsymbol{r}'-\boldsymbol{r})\,{\rm d}\boldsymbol{r}',\;\; \text{~for~} \beta = 1, 2.
\end{align*}
In the limit of a bulk hard-sphere fluid in the absence of external fields, vector densities $\boldsymbol{n}_1$ and $\boldsymbol{n}_2$ vanish.
In the same limit, the four scalar weighted densities reduce to the sum of bulk densities for all species ($n_0$) and the 1D ($n_1$), 2D ($n_2$), and 3D ($n_3$) packing fractions.
The ``weight functions" $\omega_i^{(\alpha)}$ and $\boldsymbol{\omega}_i^{(\beta)}$, characterizing the geometry of particles ({ion-ion steric correlations} with radius $R_i$ for ion species $i$), are given by \cite{Da:2014}
\begin{align}
	\omega_i^{(3)}(\boldsymbol{r})&=\theta (|\boldsymbol{r}|-R_i)\\
	\omega_i^{(2)}(\boldsymbol{r})&=|\nabla\theta(|\boldsymbol{r}|-R_i)|=\delta(|\boldsymbol{r}|-R_i)
	\label{w2}\\
	\boldsymbol{\omega}_i^{(2)}(\boldsymbol{r}) & =\nabla\theta(|\boldsymbol{r}|-R_i)=\frac{\boldsymbol{r}}{r}\delta(|\boldsymbol{r}|-R_i)
	\label{wv2}\\
	\omega_i^{(0)}(\boldsymbol{r})&=\omega_i^{(2)}(\boldsymbol{r})/(4\pi R_i^2)
	\label{w0}\\
	\omega_i^{(1)}(\boldsymbol{r})&=\omega_i^{(2)}(\boldsymbol{r})/(4\pi R_i)
	\label{w1}\\
	\boldsymbol{\omega}_i^{(1)}(\boldsymbol{r}) & = \boldsymbol{\omega}_i^{(2)}(\boldsymbol{r})/(4\pi R_i).
	\label{wv1}
\end{align}
In the preceding formula, $\theta$ is the Heaviside step function, with $\theta(x)=0$ for $x>0$ and $\theta(x) = 1$ for $x\leq 0$, and $\delta$ denotes the Dirac delta function.

The {ion-ion electrostatic interaction} term ($\FE^{\excess}_{\text{corr}}$) can be derived using the Mean Spherical Approximation \cite{Blum:1975, Hoye:1978}
\begin{multline}
	\FE^{\excess}_{\text{corr}} = \FE^{\excess}_{\text{corr}}\left[\left\{ \rho_i^{\bulk} \right\} \right]
	- kT \int_\Omega \sum_i^N c^{(1)}_i \left( \rho_i(\vec{r}) - \rho_i^\bulk \right) d \vec{r} \\
	- \frac{kT}{2} \int_\Omega \int_\Omega \sum_{i,j}^N c^{(2)}_{ij}
	\left( \rho_i(\vec{r}) - \rho_i^{\bulk} \right)
	\left( \rho_j(\vec{r}') - \rho_j^{\bulk} \right)
	d \vec{r} d \vec{r}'
\end{multline}
where $\rho_i^{\bulk}$ is the bulk concentration of ion species $i$ and the first term describes ion correlation free energy in bulk electrolyte solution in the absence of DNA. The first-order direct correlation functions are defined as
\begin{equation}
	c_i^{(1)} = -\frac{\mu_i}{kT},
\end{equation}
where $\mu_i$ is the chemical potential of ion species $i$. The second-order direct correlation functions are defined as
\begin{equation}
	c_{ij}^{(2)}\left( \vec{r} - \vec{r}' \right) =
	\begin{cases}
		-\frac{q_i q_j}{kT \epsilon} \left( \frac{2B}{\sigma_{ij}} - \left( \frac{B}{\sigma_{ij}} \right)^2 \left| \vec{r} - \vec{r}' \right| - \frac{1}{\left| \vec{r} - \vec{r}' \right|} \right) & \left| \vec{r} - \vec{r}' \right| \leq \sigma_{ij} \\
		0 & \left| \vec{r} - \vec{r}' \right| > \sigma_{ij},
	\end{cases}
\end{equation}
where $q_i$ is the charge of ion species $i$, $\epsilon$ is the dielectric constant of the solvent, $\sigma_{ij} = \left( \sigma_i + \sigma_j \right)/2$ is the hard-sphere contact distance between ions of diameters $\sigma_i$ and $\sigma_j$, $B$ is given by
\begin{equation}
	B = \frac{1}{\xi} \left( \xi + 1 - \sqrt{1 + 2 \xi} \right),
\end{equation}
$\xi = \kappa \sigma_{ij}$, $\kappa$ is the inverse Debye length $\kappa^2 = l_B \sum_i q_i^2 \rho_i^{\bulk}$, $l_B = \frac{e^2}{kT\epsilon}$ is the Bjerrum length, and $e$ is the unit charge.
The direct Coulomb free energy term can be calculated exactly
\begin{equation}
	\FE^{\excess}_{\text{C}} = \frac{kT l_B}{2} \int_\Omega \int_\Omega \sum_{i,j}^N \frac{q_i q_j}{\left| \vec{r} - \vec{r}' \right|} \rho_i(\vec{r}) \rho_j(\vec{r}') d \vec{r} d \vec{r}'.
\end{equation}
Finally, the {ion-solvent interaction} term $\FE^{\excess}_{\text{solv}}$ models ion-water interactions with a square well potential
\begin{equation}
	V(\vec{r} - \vec{r}') = \begin{cases}
		\infty & \left| \vec{r} - \vec{r}' \right| < \sigma_{ij} \\
		-\varepsilon & \sigma_{ij} \leq \left| \vec{r} - \vec{r}' \right| \leq \sigma_{ij} + h \\
		0 & \sigma_{ij} + h < \left| \vec{r} - \vec{r}' \right|,
	\end{cases}
\end{equation}
where $\varepsilon$ is the well depth, and $h$ is the well width.
For the current study, $h = 0.2 \varsigma$ is the well width for interactions between ions and water and $\varsigma$ is the sum of radii of interacting particles \cite{Cao:2005}.
The following well depths were calculated using SPC/E water using the parameters from Horinek et al:  $\varepsilon_{\Sr} = 0.01038$ eV, $\varepsilon_{\Cl} = 0.0053894$ eV, $\varepsilon_{\Rb} = \epsilon_{\CoHex} = 0.0021$ eV \cite{Horinek:2009}.
Simulations of concentration dependence of ion activity coefficients in RbCl and CoHex$\Cl_3$ solutions demonstrated that adding attractive ion-water interactions does not affect the ion chemical potential (see Supporting Information).

Minimization of the excess free energy functional $\FE^{\excess}$ with respect to the water and ion densities gives
\begin{equation}
	\rho_i(\vec{r}) = \exp \left( \frac{\mu_i}{kT} - \frac{1}{kT} \frac{\delta \FE^{\excess}}{\delta \rho_i(\vec{r})} \right).
\end{equation}
We solve Poisson's equation
\begin{equation}
	-\nabla \cdot \epsilon(\vec{r}) \nabla \varphi(\vec{r}) = \sum_i q_i \rho_i(\vec{r})
	\label{eqn:poisson}
\end{equation}
for the electrostatic potential ($\varphi(\vec{r})$) where $\epsilon(\vec{r})$ is the dielectric coefficient.
For an infinitely long uniformly charged cylinder in electroneutral conditions, the potential
\begin{equation}
	\varphi(r) = \frac{4 \pi}{\epsilon} \int_r^\infty t \log \left( \frac{r}{t} \right) \sum_i^N q_i \rho_i(t) dt.
\end{equation}
Using this potential for the cylinder model and a numerical solution to Poisson's equation (Eq.\ \ref{eqn:poisson}) for the 3D DNA model, the expression for the densities is
\begin{eqnarray}
	\rho_i(\vec{r}) &=& \exp \left( \frac{\mu_i}{kT} - \frac{q_i \varphi(\vec{r})}{kT}- \frac{1}{kT} \frac{\delta \left( \FE^{\excess}_{\text{hs}} + \FE^{\excess}_{\text{corr}} + \FE^{\excess}_{\text{solv}} \right)}{\delta \rho_i(\vec{r})} \right).
	\label{eqn:new-dens}
\end{eqnarray}
The resulting system of Eqs.\ \ref{eqn:poisson} and \ref{eqn:new-dens} was solved iteratively to self-consistency using the numerical procedure described {in detail} by Meng \cite{Da:2014}.
In particular, equilibrium ion density distributions were obtained using a relaxed Gummel iterative procedure for 3D systems and Picard iterations in 1D.
Convergence was considered to be achieved when the maximum difference between the input and the output density profiles between iterations was smaller than $10^{-6}$.
{The solution of Eqs.\ \ref{eqn:poisson} and \ref{eqn:new-dens} encompasses the equilibrium distribution of the densities of all ion species, corresponding to the minimum of the total free energy; the corresponding free energies for each contribution; and the chemical potentials.
``Panoramic'' density distributions representing angular distributions of ions on DNA backbone and in minor grooves were calculated along the corresponding helical shells.
For each, angle the ion densities were averaged within the shells over $r$ and $z$.
For ions on the DNA backbone and in the minor grooves, the radial positions of the shells was defined as $1 < r < (1 + \sigma)$ nm and $0.5 < r < 1$ nm, respectively.}

Three main features distinguish our approach from previous cDFT models \cite{Goel:2008, Goel:2011, Patra:1999}.
First, our model includes a full representation of the coarse-grained DNA topology and a discrete distribution of charges.
Second, we use Pauling diameters for ions and van der Waals diameters for water molecules as opposed to previous restricted models where all species have the same diameter.
Finally, our model includes water-ion attractive interactions.

\subsection*{Anomalous small-angle X-ray scattering curve calculations}
ASAXS profiles were calculated using the ion density distributions $\rho_i(r)$ around DNA.
In the 3D model, ion densities were averaged in cylindrical coordinates over the cylinder azimuthal angle $\phi$ and length $z$ for each radial distance $r$ from the DNA axis.
The excess form factor for ion species $\alpha$ was calculated as
\begin{equation}
	F_{\text{ion}, \alpha}(Q) = a_\alpha \int \rho_\alpha(r) e^{-\imath Q r} dr,
\end{equation}
where $a_\alpha$ is a constant related to the average electron density of ion species $\alpha$ and $Q$ is the scattering vector.
In the current study, we only consider the excess form factor due to cation species; the chloride anion has no ASAXS response.
Furthermore, we only consider a single cation species at a time so that $F_{\text{ion}}(Q) = F_{\text{ion}, \alpha}(Q)$.

The excess form factor of DNA ($F_{\text{DNA}}(Q)$) was calculated using AquaSAXS \cite{Poitevin:2011}.
from the form factor of DNA in vacuo ($F_{\text{DNA}}^{\text{vac}}(Q)$), the form factor of the volume of water excluded by DNA ($F_{\text{DNA}}^{\text{excl}}(Q)$), and the form factor of hydration shell of the DNA ($F_{\text{hsh}}(Q)$):
\begin{equation}
	F_{\text{DNA}}(Q) = F_{\text{DNA}}^{\text{vac}}(Q) - \rho_w F_{\text{DNA}}^{\text{excl}}(Q) + \rho_w F_{\text{hsh}}(Q),
\end{equation}
where $\rho_w$ is the bulk density of water.
The form factor of the hydration shell is calculated using water density maps, $\rho_{\text{hsh}}(r)$, obtained via AquaSol \cite{Koehl:2010}, which employs the Poisson-Boltzmann formalism with water treated as an assembly of self-oriented dipoles:
\begin{equation}
	F_{\text{hsh}}(Q) = b \int \left( \frac{\rho_{\text{hsh}}(r)}{\rho_w} - 1 \right) e^{-\imath Q r} dr
\end{equation}
where $b$ is a scale factor to adjust the hydration shell contribution (usually $b=1.0$) and integration is performed over the region where solvent density deviates from the bulk by a factor larger (in magnitude) than $\pm 10^{-4}$.

The ASAXS intensity is then calculated from these quantities as
\begin{equation}
	I(Q) = 2\left( f'_{\text{ion}}(E_{1})-f'_{\text{ion}}(E_{2}) \right) \left(f_{\text{DNA}}N_{\text{ion}}F_{\text{DNA}}(Q)F_{\text{ion}}(Q)
	+
	f_{\text{ion0}}N^{2}_{\text{ion}}F_{\text{ion}}(Q)^{2} \right) + \left(f'^{2}_{\text{ion}}(E_{1}) - f'^{2}_{\text{ion}}(E_{2})\right) N^{2}_{\text{ion}}F^{2}_{\text{ion}}(Q)
\end{equation}
where $f'_{\text{ion}}(E_{i})$ is the energy-dependent real part of ion anomalous scattering factor,$E_1$ is the energy far from the X-ray absorption edge of the ion, $E_2$ is the energy near the edge where ion scattering is suppressed by absorption, $f_{\text{ion0}}$ is the energy independent solvent-corrected scattering factor, $f_{\text{DNA}}$ is the effective number of electrons from DNA and $N_{\text{ion}}$ is the number of excess ions around DNA \cite{Pabit:2010} (see Supporting Information for more details).
Since experimental data are available in arbitrary units, theoretical intensities were uniformly scaled with a common scaling factor, chosen to match the experimental and calculated intensities, obtained using 3D cDFT-full model, at low $Q$.

\section*{Results}

{Tables \ref{tab:models} and \ref{tab:results} provide qualitative and quantitative results on the performance of the models.
These results are described in greater detail below.}

\begin{table}
	\caption{{Numbers of condensed ions for the models described in Table \ref{tab:models} and experimental results \cite{Pabit:2010} (where available).}}
	\label{tab:results}
	\centering
	\begin{tabular}{p{1.5in}|p{0.6in}p{0.6in}p{0.6in}}
		\toprule
		Model 					& Rb$^+$	& Sr$^{2+}$	& CoHex$^{3+}$	\\
		\hline
		NLPB 					& 23.0 		& 13.6 		& 5.6			\\
		cDFT, no correlation	& 23.4		& 13.7		& 5.6			\\
		cDFT, no ion solvation	& 34.9		& 25.1		& 5.7			\\
		cDFT, full model		& 34.6		& 16.6		& 5.7			\\
		Experiment 				& 34 $\pm$ 3& 19 $\pm$ 2& No data		\\
		\bottomrule
	\end{tabular}
\end{table}

\subsection*{Comparison between DNA Model systems}
\begin{figure}[t]
	\centering
	(a)~\includegraphics[keepaspectratio,width=0.45\linewidth]{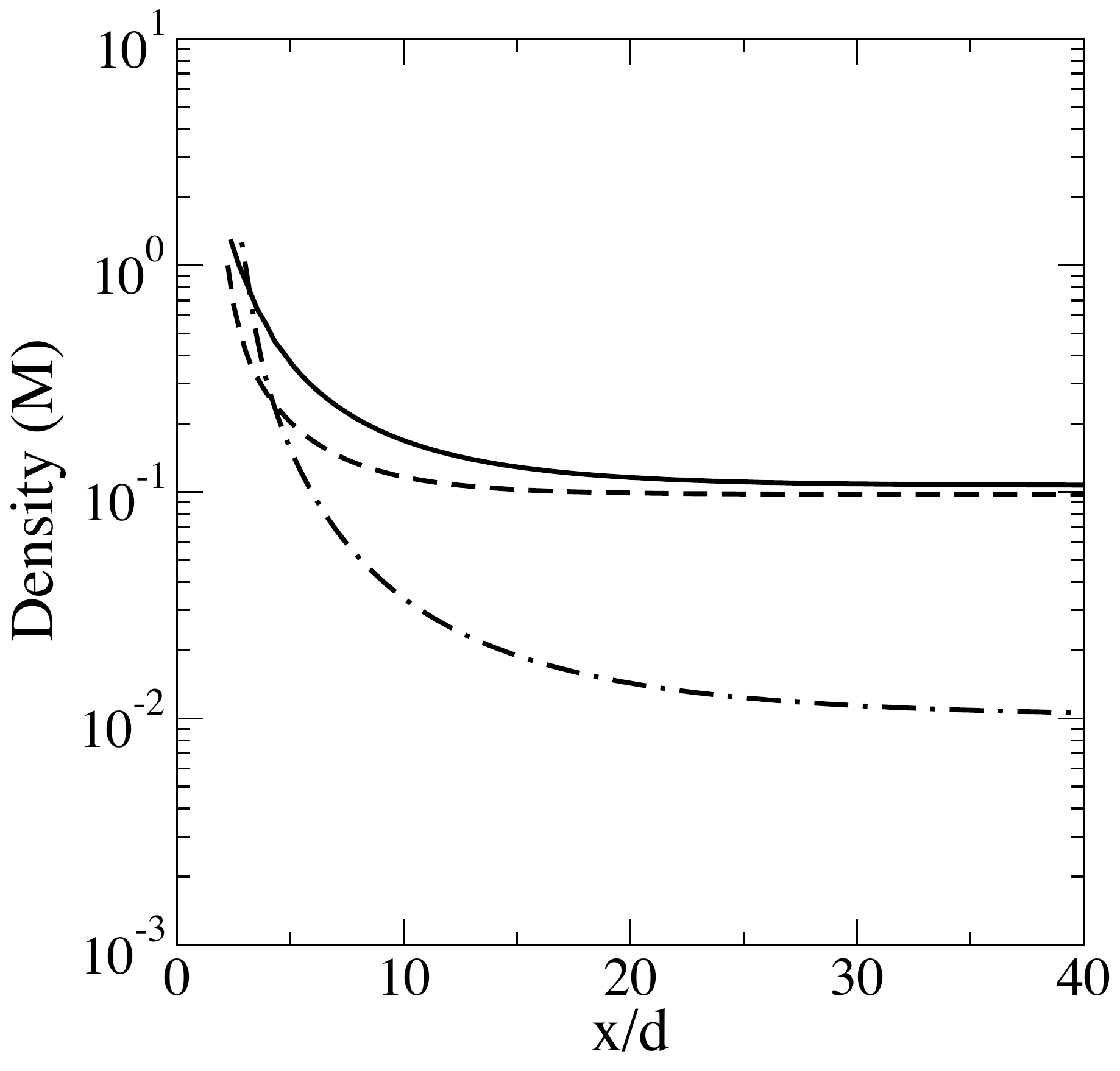}
	(b)~\includegraphics[keepaspectratio,width=0.45\linewidth]{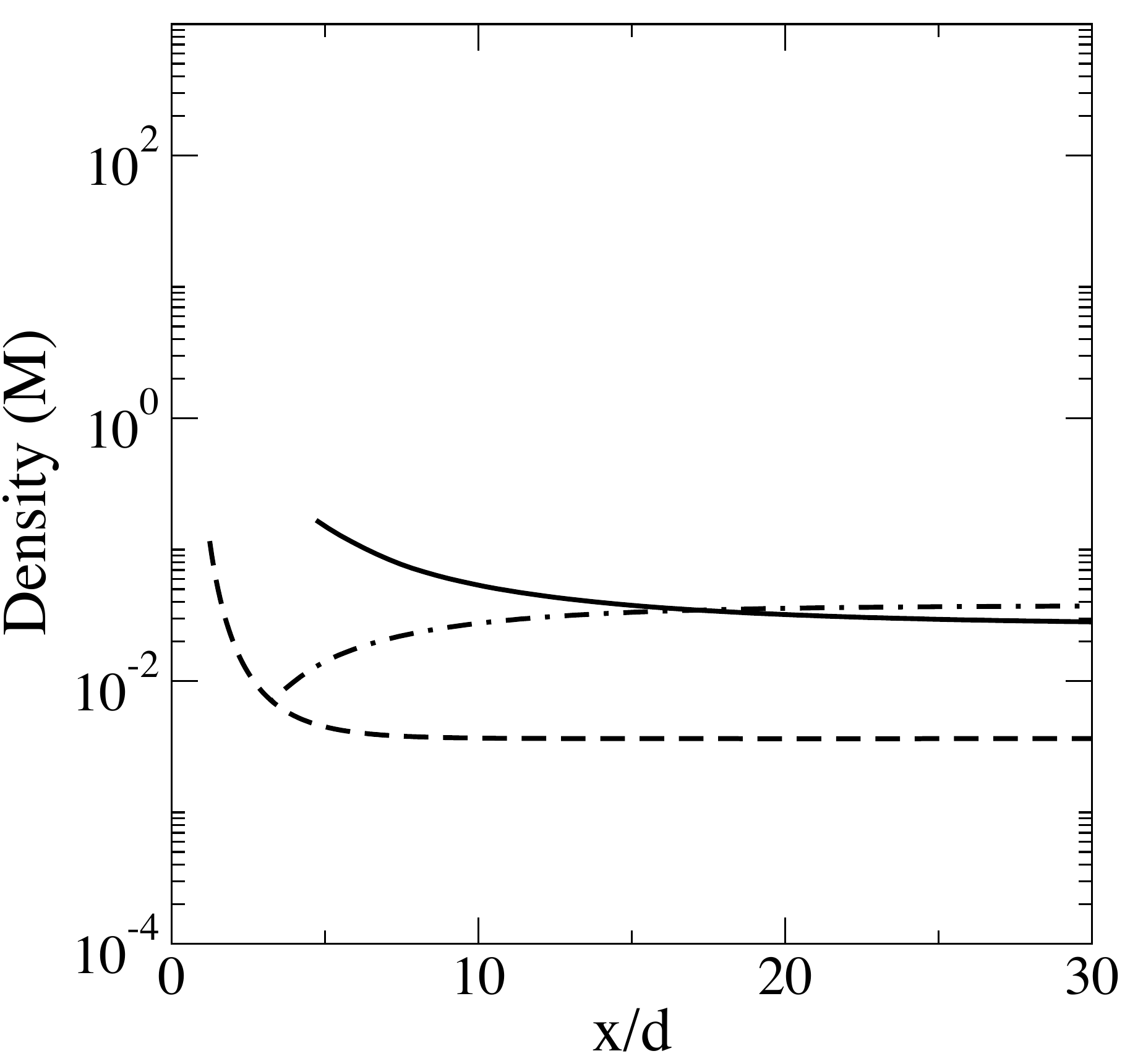}
	\caption{Ionic distributions around a uniformly charged cylinder.
	{The distance $x/d$ is the distance ($x$) from the cylinder surface scaled by the ion radius ($d$).}
	(a) Solutions of 100 mM NaCl, 100 mM RbCl, and 10 mM SrCl$_2$ in a 1 mM NaCl buffer; concentration profiles are shown for Na$^+$ (solid line), Rb$^+$ (dashed line), and Sr$^{2+}$ (dot-dashed line).
	(b) Solutions of 5 mM CoHexCl$_3$ in a 20 mM NaCl buffer; concentration profiles are shown for Na$^+$ (solid line), CoHex$^{3+}$ (dashed line), and Cl$^-$ (dot-dashed line).}
	\label{fig:cdft-cylinder}
\end{figure}

The uniformly charged cylinder model (Fig.\ \ref{fig:dna-models} left) represents a one-dimensional case for which ionic distribution is only a function of the radial distance from the cylinder axis.
\revision{This 1D model produces monotonically decreasing  with the distance from the cylinder surface density distributions of monovalent and multivalent counterions (Fig.\ \ref{fig:cdft-cylinder}).
Competitive cation condensation in mixed 5 mM CoHexCl$_3$ + 20 mM NaCl solutions results in preferential CoHex$^{3+}$ condensation on the cylinder surface: sodium ions are not found in the immediate vicinity of the DNA (Fig.\ \ref{fig:cdft-cylinder}b).
This competition is in qualitative agreement with experimental observations of a negligible effect of Na$^+$ on CoHex$^{3+}$ binding when NaCl concentration is below 40 mM \cite{Braunlin1987}.}

\revision{For monovalent ions, the 1D cDFT calculations predict 91.5\% and 77.5\% DNA charge neutralization by Na$^+$ and Rb$^+$, respectively.
Such differences in monovalent cation condensation on DNA were not observed experimentally \cite{Andresen:2004}, demonstrating a fundamental deficiency of a uniformly charged cylinder model for simulating ionic atmosphere around DNA.
For divalent ions, the 1D cDFT calculations predict charge inversion at the DNA surface in SrCl$_2$ solution.
Note that charge inversion in the presence of multivalent salts has also been observed in cDFT and MC simulations for a cylinder DNA model \cite{Goel:2008, Goel:2011}.
However, we do not see this effect in our more detailed 3D DNA geometry simulations (see below).
Finally, for trivalent ions, 90\% DNA charge neutralization is found within 5 CoHex radii from the cylinder surface or within the region where excess CoHex concentration is present (Fig.\ \ref{fig:cdft-cylinder}b).}

\begin{figure}[t]
	\centering
	(a)~\includegraphics[keepaspectratio,width=0.45\linewidth]{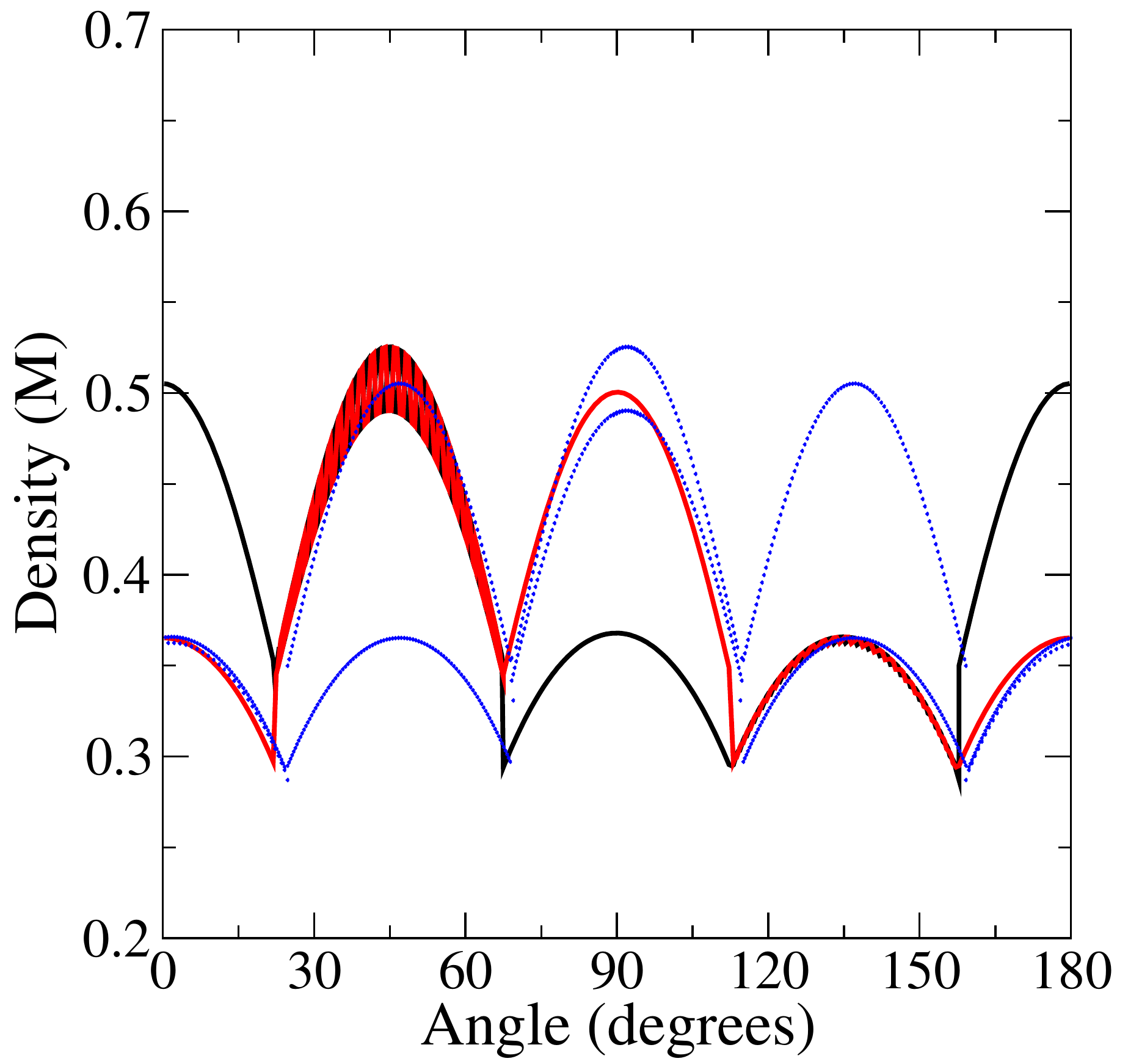}
	(b)~\includegraphics[keepaspectratio,width=0.45\linewidth]{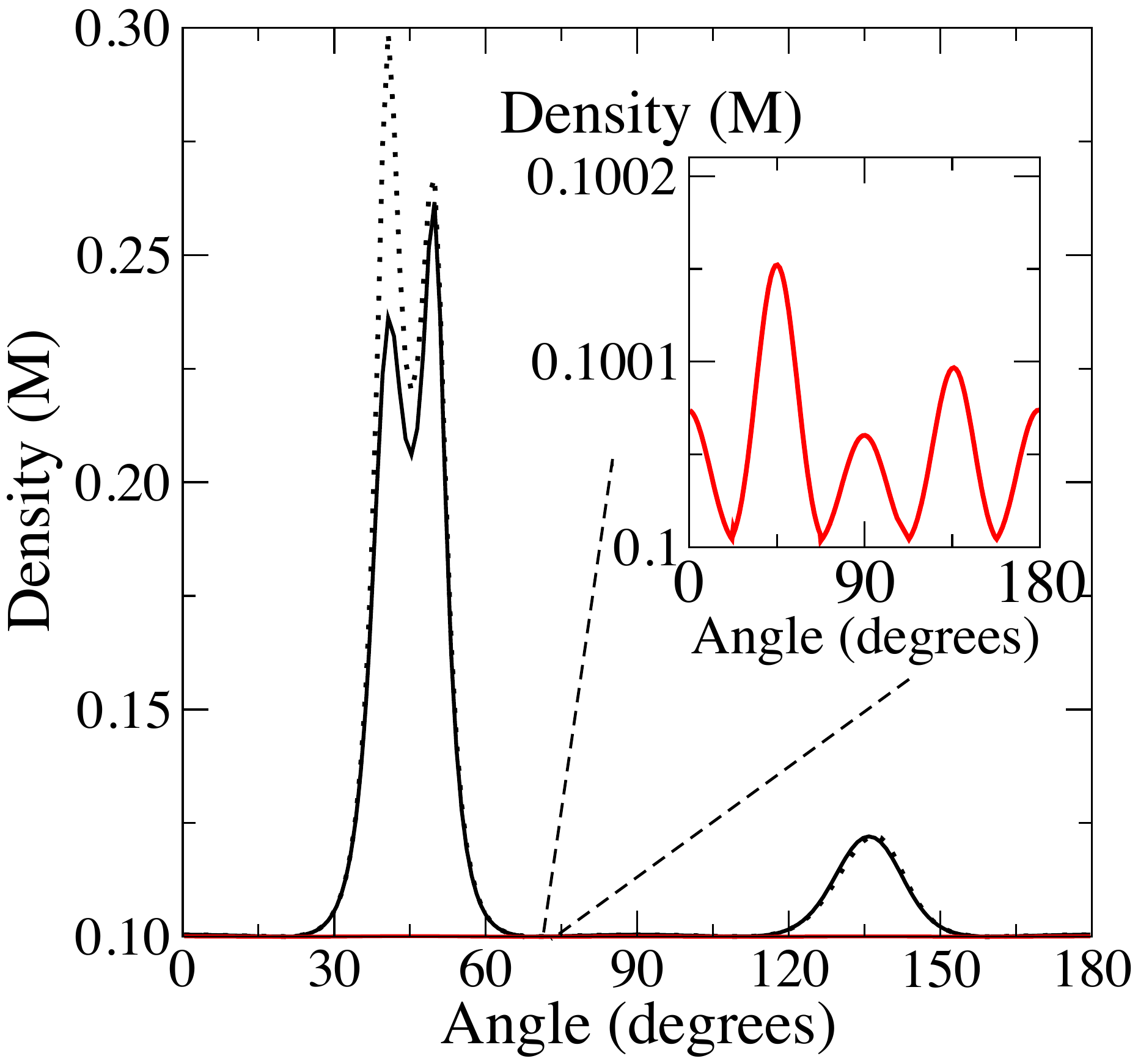}
	(c)~\includegraphics[keepaspectratio,width=0.45\linewidth]{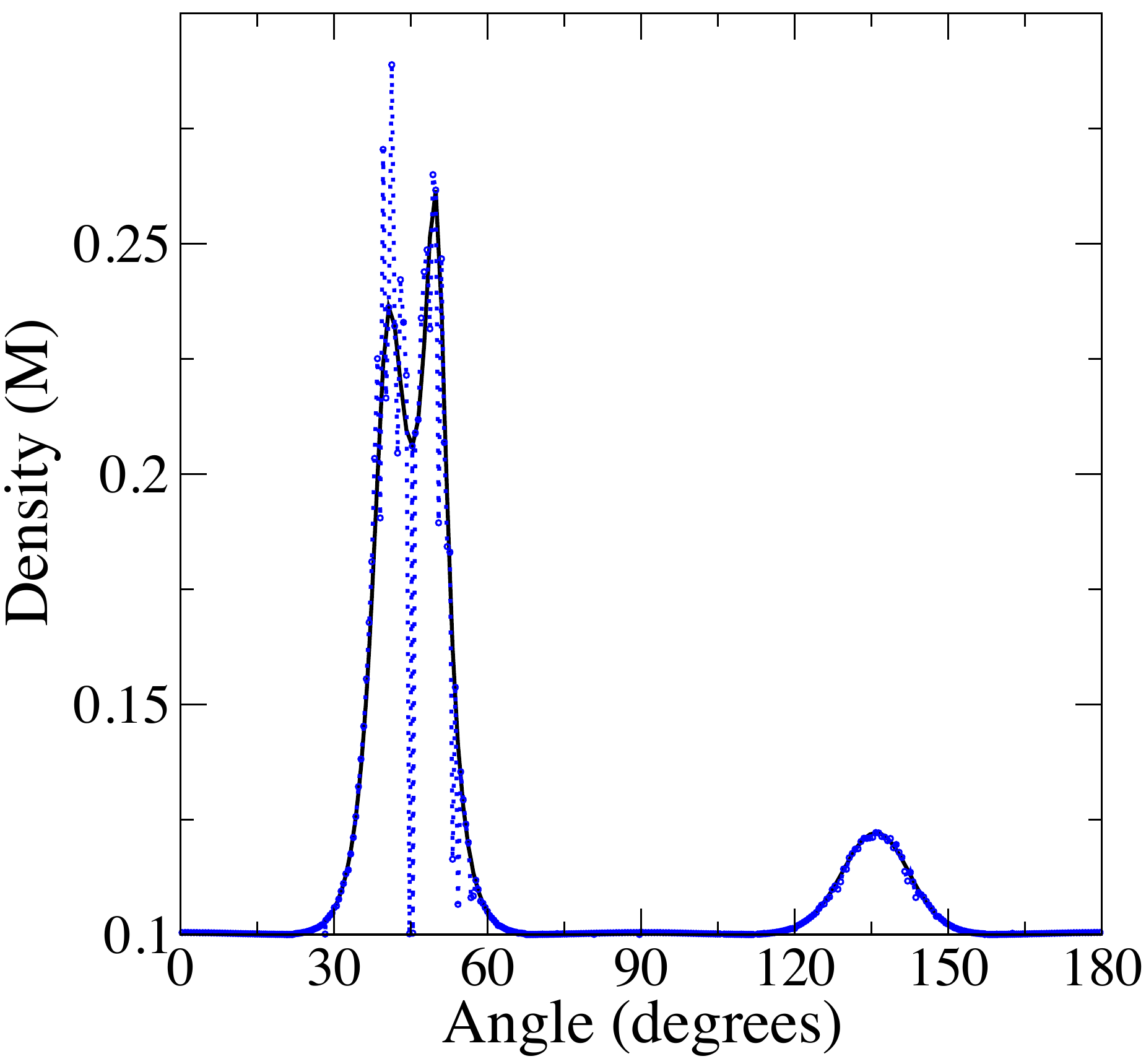}
	(d)~\includegraphics[keepaspectratio,width=0.45\linewidth]{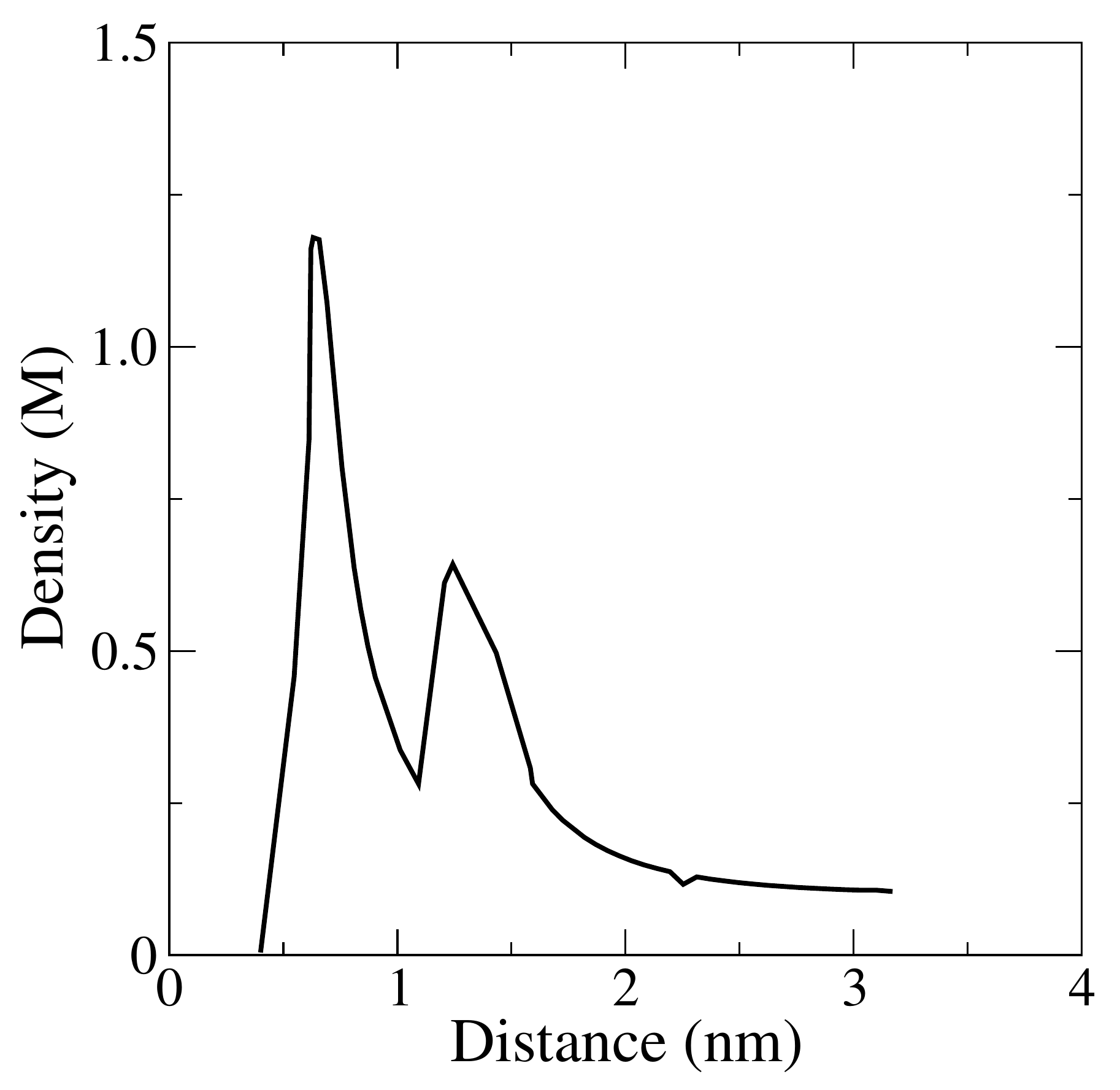}
	\caption{\revision{Cation distributions around DNA calculated using (a) cDFT with the full model, (b) cDFT with no ion-correlation interactions (cDFT-nc, dotted line), and (b,c) NLPB.
	In panels (a-c), panoramic Rb$^+$ density distributions are shown on the DNA backbone (black lines) and in the minor groove (red lines) as defined in the manuscript text.
	The inset in panel (b) shows a zoom-in into a low-density region.
	Panoramic views of cation distributions around DNA in 100 mM NaCl are shown in blue in panels (a) and (c) for comparison.
	The radial Rb$^+$ density distribution calculated from the full cDFT model is shown in panel (d).}}
	\label{fig:cdft-Rb-helical}
\end{figure}

\begin{figure}[t]
	\centering
	(a)~\includegraphics[keepaspectratio,width=0.45\linewidth]{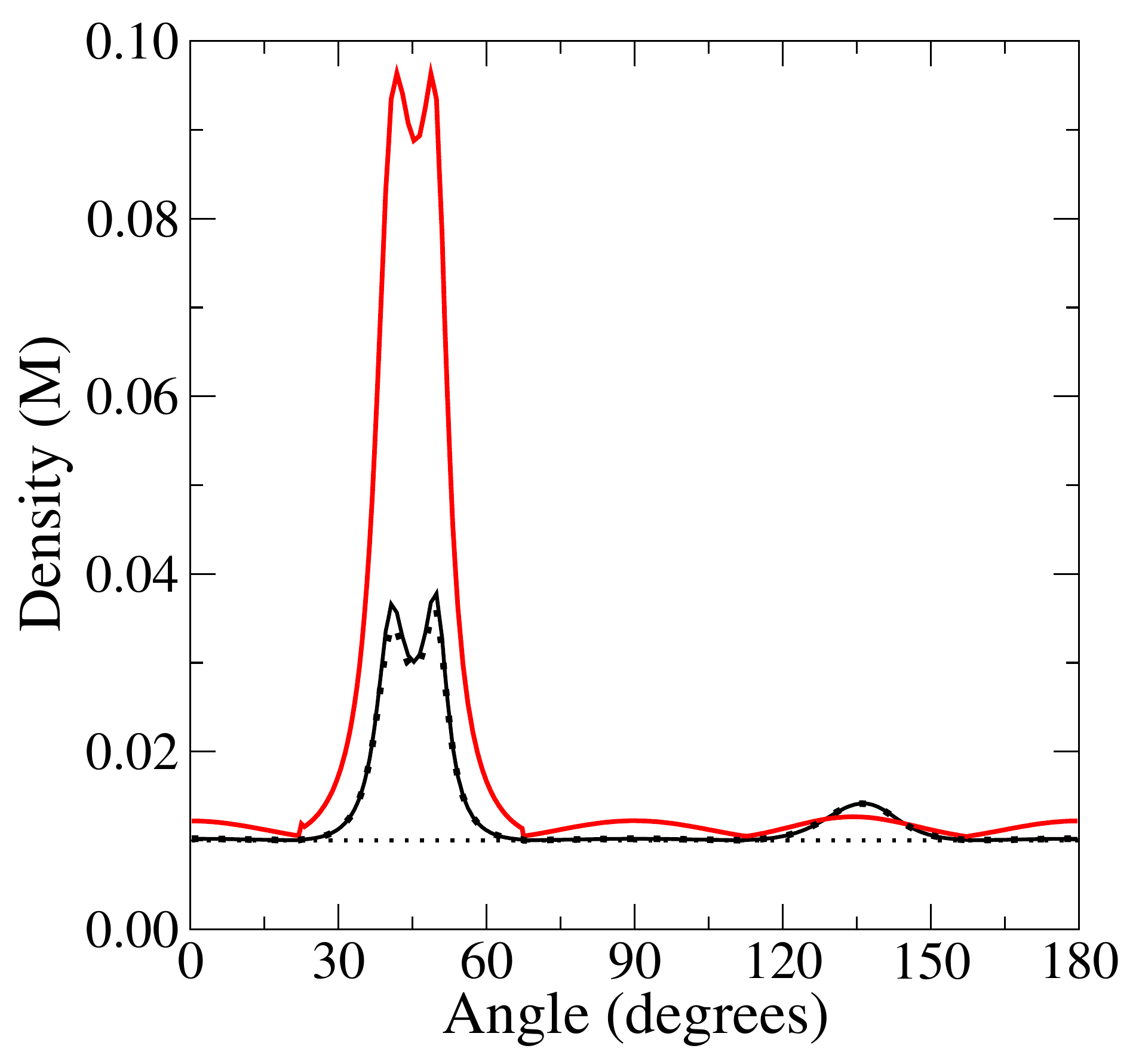}
	(b)~\includegraphics[keepaspectratio,width=0.45\linewidth]{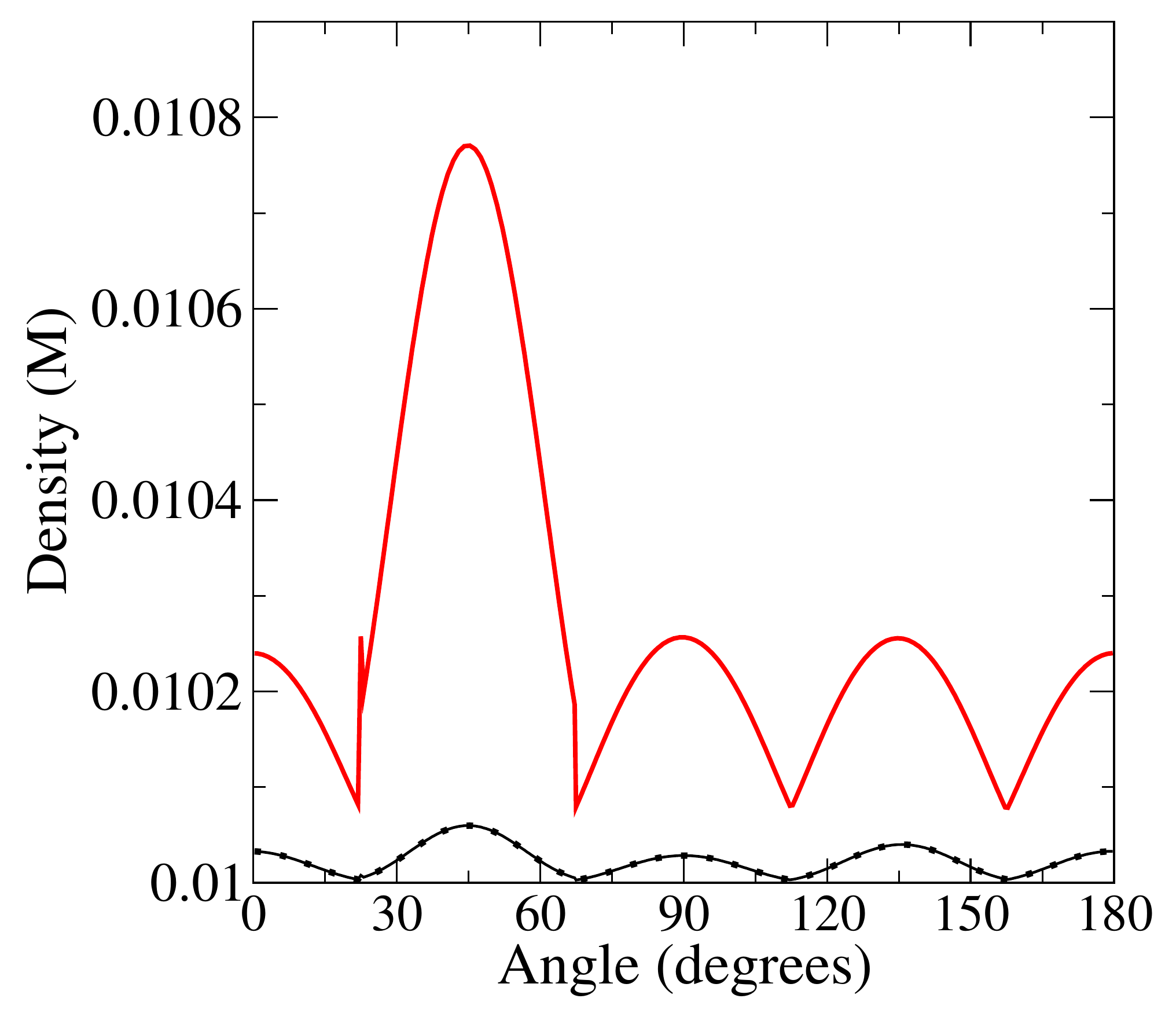}
	(c)~\includegraphics[keepaspectratio,width=0.45\linewidth]{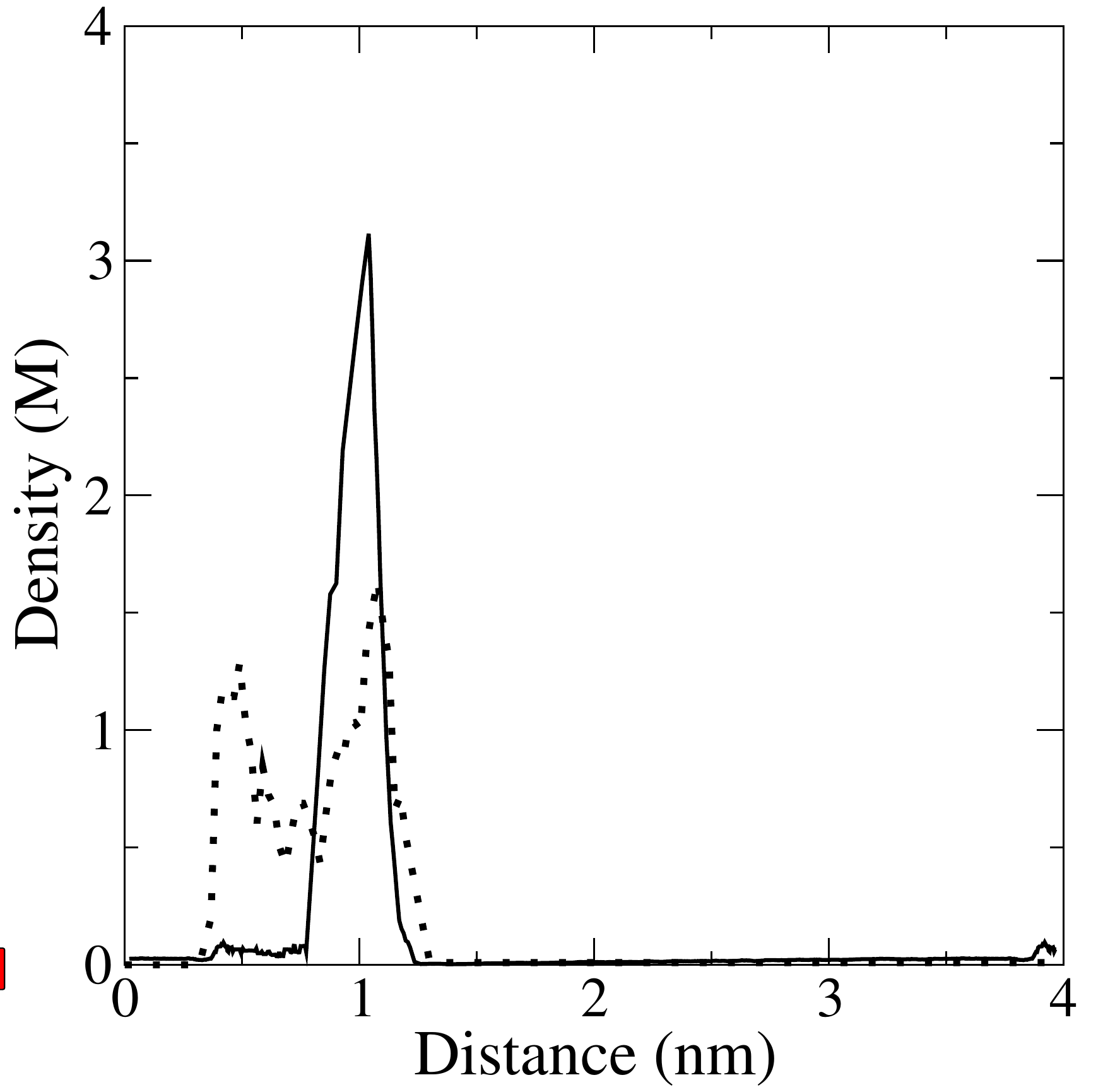}
	\caption{\revision{Panoramic density distributions of Sr$^{2+}$ ions on (a) the DNA {backbone} and in (b) the DNA minor grooves obtained 10 mM SrCl$_2$.
	Backbone and minor groove definitions are provided in the text.
	Results are shown for the full cDFT model (solid red line), cDFT with no corrlations (cDFT-nc, dotted line), and the NLPB model (solid black line).
	Panel (c) shows radial densities of Sr$^{2+}$ ions around a DNA molecule calculated using the full cDFT model (solid line) and the cDFT model without ion-water attractive interactions (dotted line).}}
	\label{fig:cdft-Sr-helical}
\end{figure}

\begin{figure}[t]
	\centering
	(a)~\includegraphics[keepaspectratio,width=0.45\linewidth]{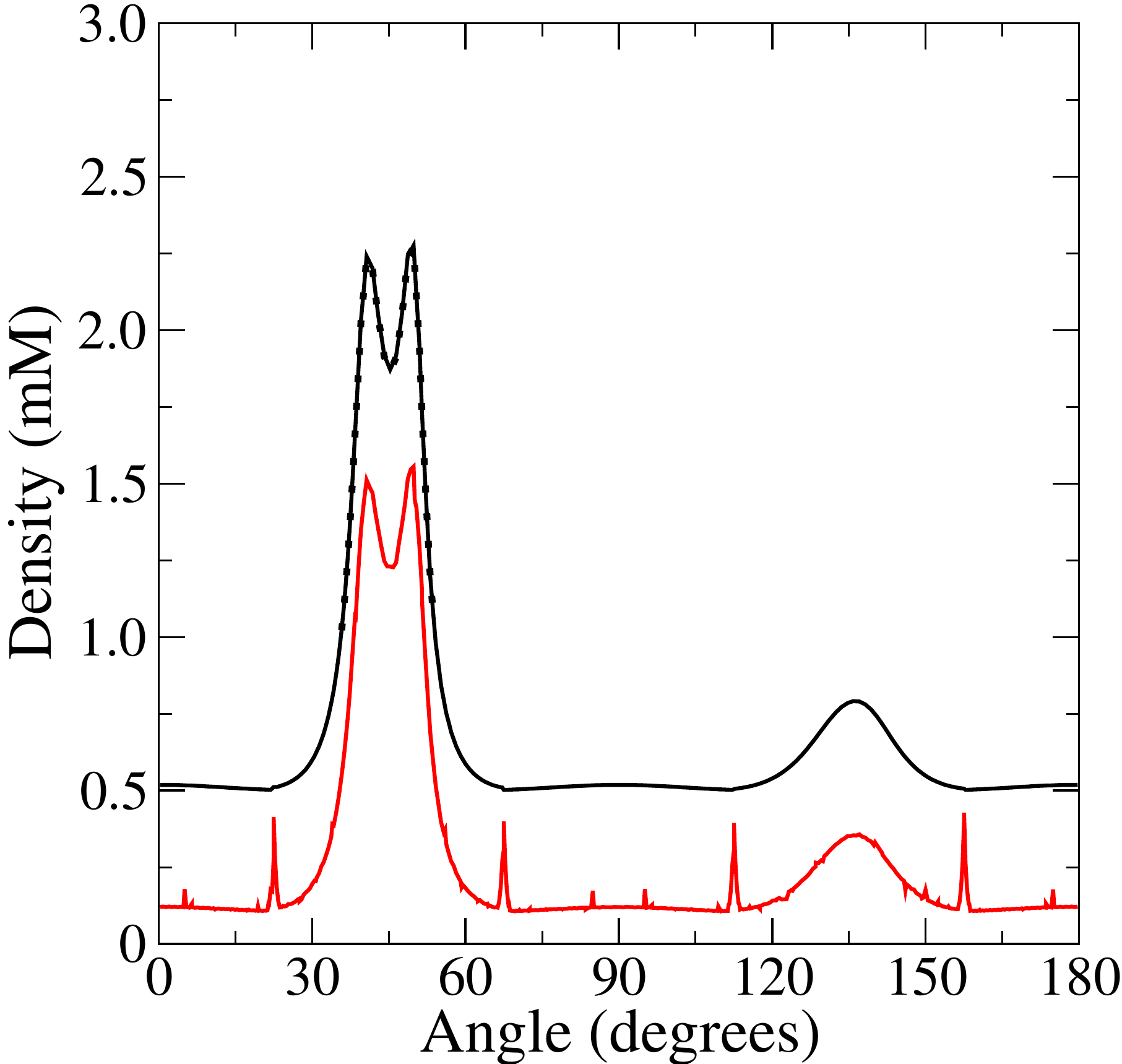}
	(b)~\includegraphics[keepaspectratio,width=0.45\linewidth]{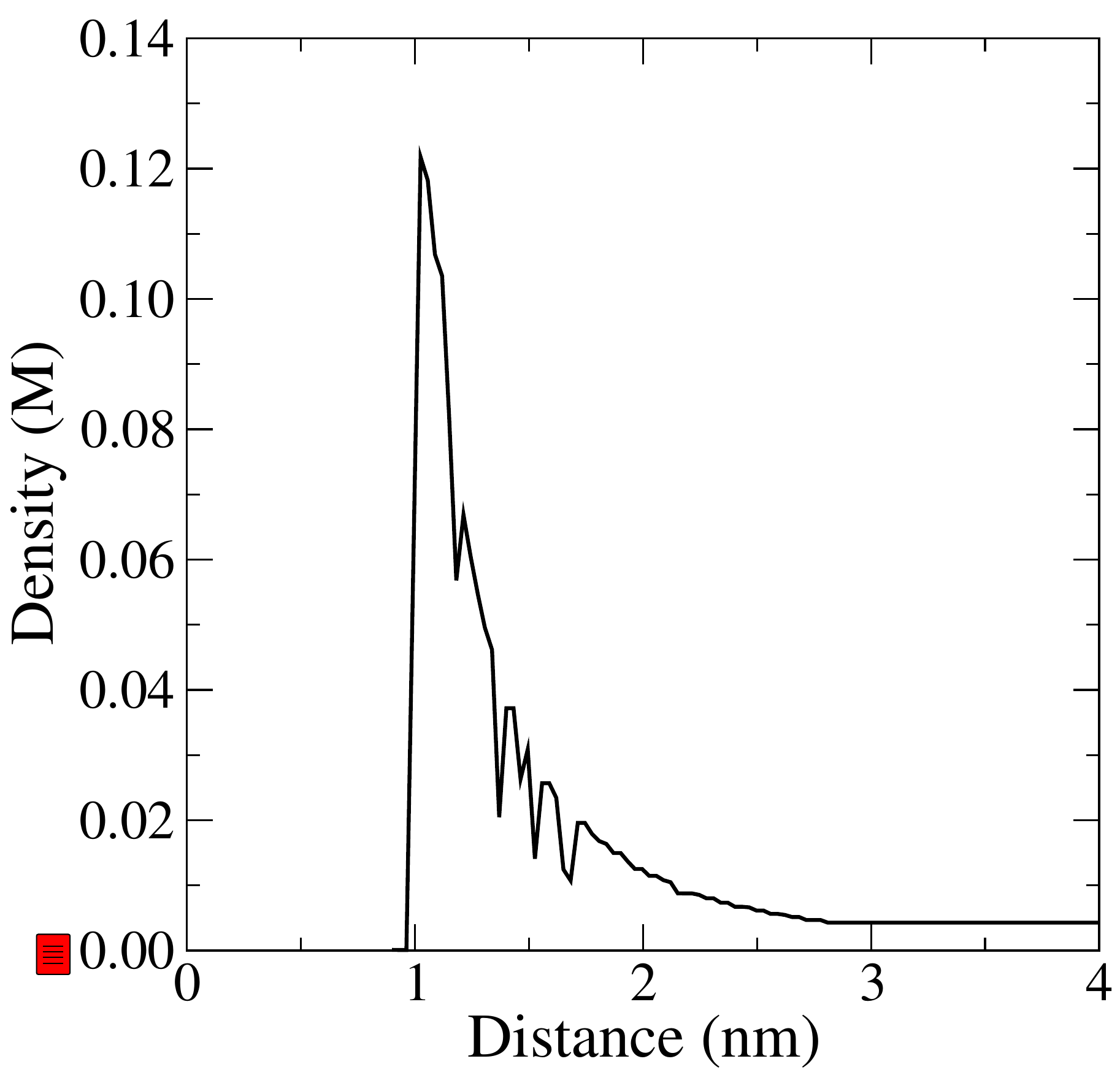}
	\caption{(a) Panoramic density distributions of CoHex$^{3+}$ ions on DNA {backbone} obtained using \revision{the full cDFT (solid red line) model, the cDFT model with no correlations (cDFT-nc, dotted line), and the NLPB model (solid black line) for 0.5 mM CoHexCl$_3$.
	The NLPB and cDFT-nc curves have been shifted by 0.5 mM for clarity.
	(b) Radial {density} distributions of CoHex$^{3+}$ ions around the DNA molecule.
	Note: as seen in panel (b), there is zero CoHex density$^{3+}$ in the minor groove, so the corresponding panoramic density is not shown.}}
	\label{fig:cdft-CoHex-helical}
\end{figure}

We also performed 3D cDFT calculations of the same electrolyte solutions surrounding the helical discrete charge model (Fig.\ \ref{fig:dna-models} right).
Fig.~\ref{fig:cdft-Rb-helical} shows cDFT results for the monovalent ion Rb$^+$.
As shown in the panoramic density profiles, cDFT predicts a two-peak radial density distribution of Rb$^+$ {(Fig.~\ref{fig:cdft-Rb-helical})}: first peak at around 0.6 nm is due to cation penetration into DNA minor grooves and the second peak at 1.2 nm to Rb$^+$ condensation on DNA {backbones}.
These data are in good quantitative agreement with molecular dynamics results obtained using TIP3P water model \cite{Giambasu:2014}.
Fig.\ \ref{fig:cdft-Sr-helical} shows cDFT results for the divalent ion Sr$^{2+}$.
In the case of Sr$^{2+}$, the effect of {ion-solvent interactions} can be clearly seen in the density distribution of Sr$^{2+}$ with respect to the DNA axis {(Fig.~\ref{fig:cdft-Sr-helical})}.
While both cDFT models - with and without {ion-solvent interactions} - produce two-peak Sr$^{2+}$ density distributions at the same positions with respect to the DNA axis, the density distributions are qualitatively different.
In particular, the model without {ion-solvent interactions} predicts much higher Sr$^{2+}$ concentration in the DNA grooves than on {the backbone} , while the trend is reversed in the model with {ion-solvent interactions}.
The 3D cDFT results for trivalent CoHexCl$_3$ solutions are shown in Fig.\ \ref{fig:cdft-CoHex-helical}b and are very similar to those obtained from the 1D cDFT model.

\begin{figure}[t]
	\centering
	(a)~\includegraphics[keepaspectratio,width=0.45\linewidth]{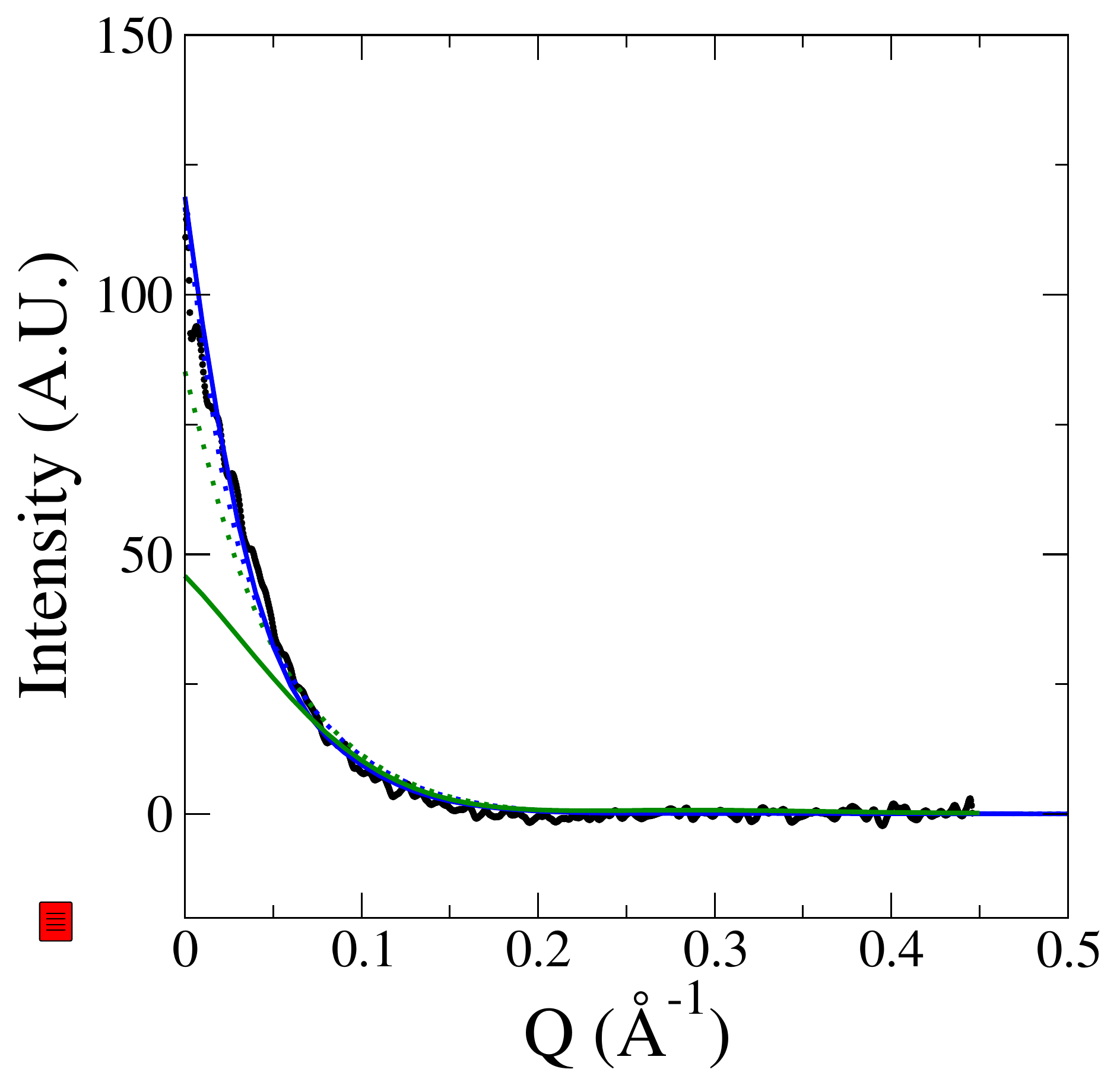}
	(b)~\includegraphics[keepaspectratio,width=0.45\linewidth]{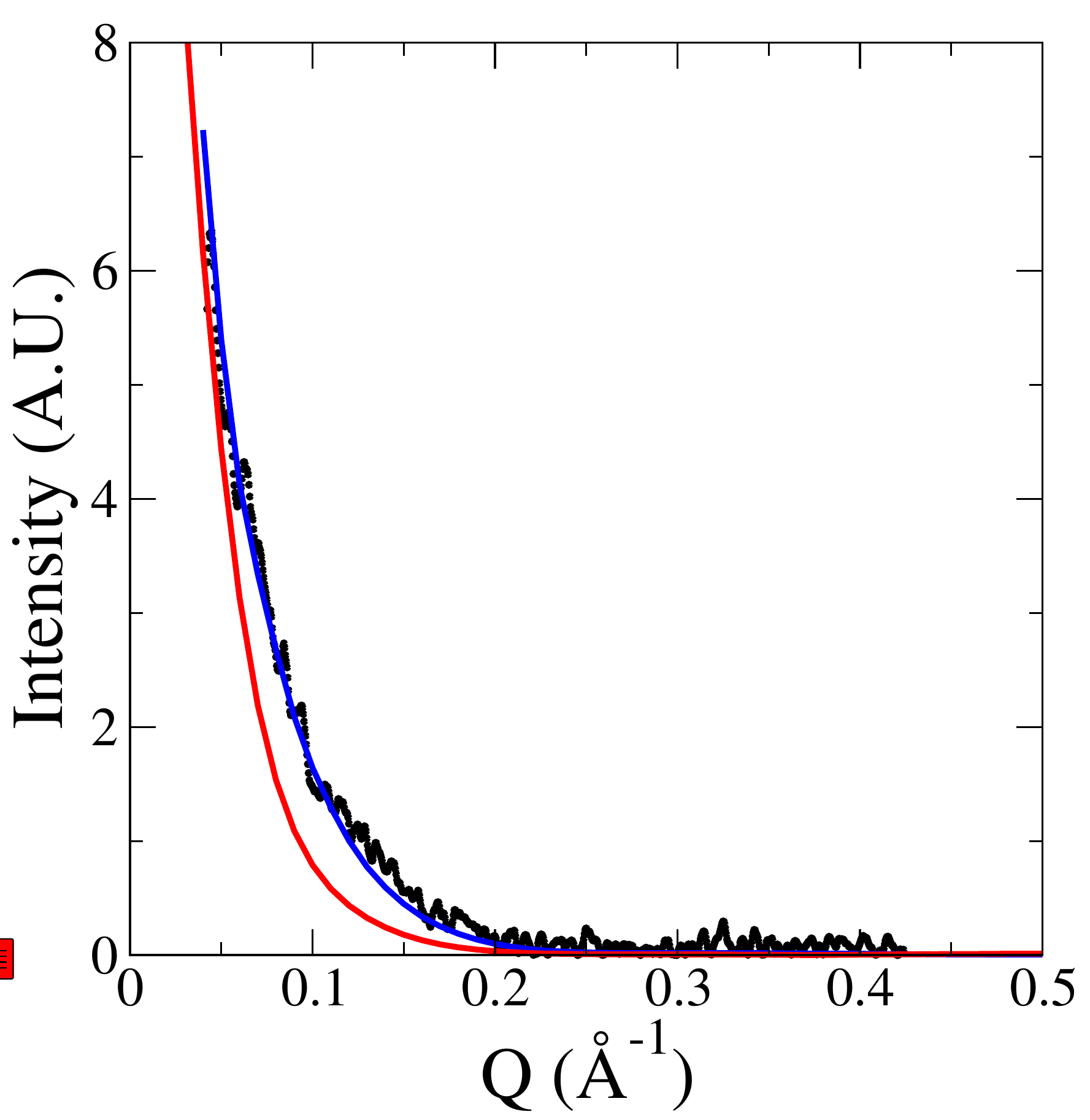}
	\caption{Simulated and experimental ASAXS profiles for 25 bp DNA in (a) 100 mM RbCl and (b) 10 mM SrCl$_2$ solutions.
	\revision{Experimental data \cite{Andresen:2008,Pabit:2010} are shown as black dots.
	This figure shows simulation results using the full 3D cDFT model (blue lines), the 3D cDFT model without ion-water interactions (red lines), the full 1D cDFT model (blue dots), and the NLPB model (green lines and dots).}}
	\label{fig:rb-sr-asaxs}
\end{figure}

\begin{figure}[t]
	\centering
	\includegraphics[keepaspectratio,width=0.45\linewidth]{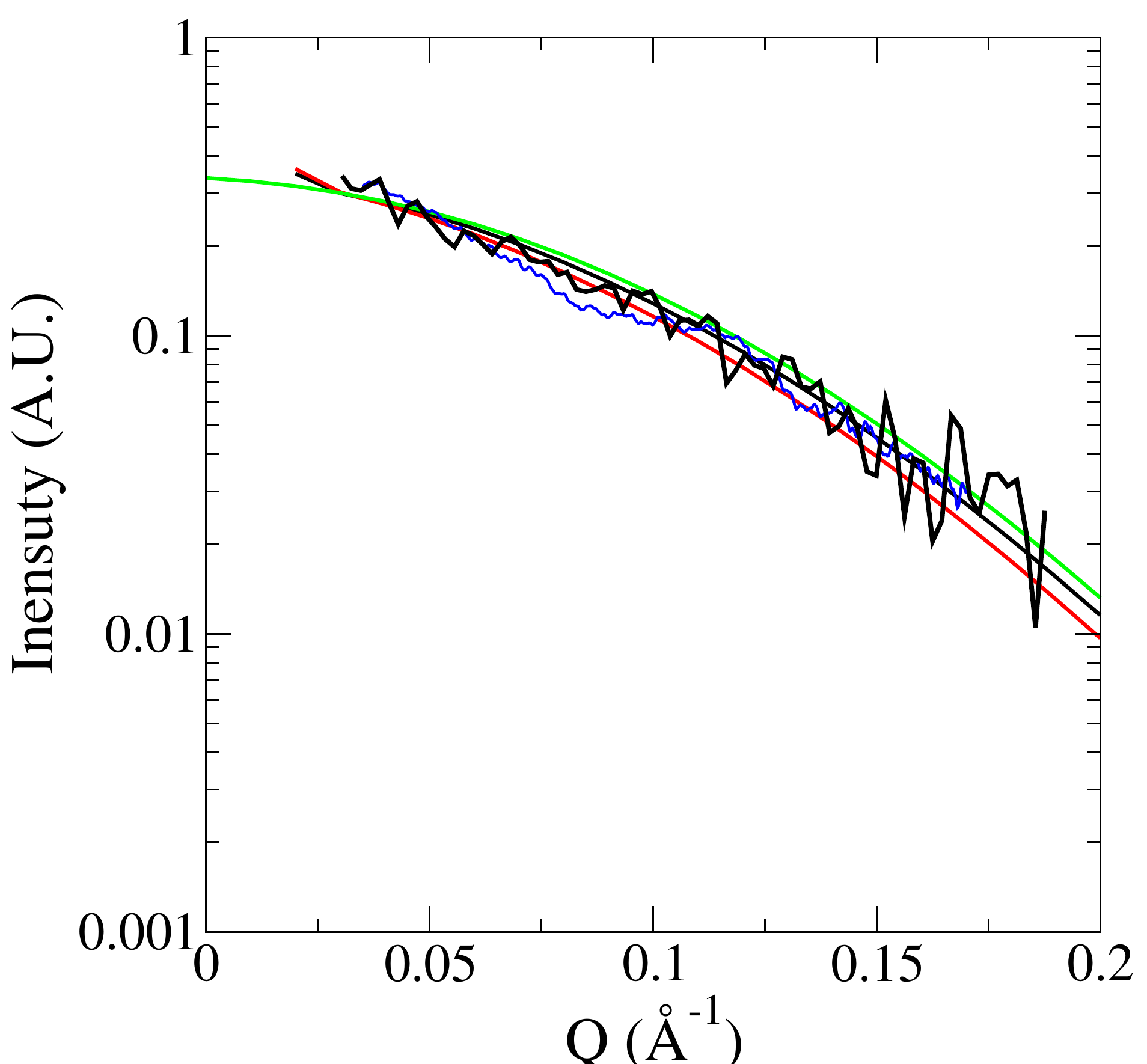}
	\caption{\revision{Simulated and experimental ASAXS profiles for 25 bp DNA in 0.5 mM CoHexCl$_3$ DNA solutions.
	Experimental data (unpublished) are shown with a thick black line and those from Andresen et al \cite{Andresen:2008} with a blue line.
	1D cDFT results are shown with a thin black line and 1D NLPB results with a red line.
	The 3D cDFT and NLPB data coincide and are shown with a green line.}}
	\label{fig:cohex-asaxs}
\end{figure}

Finally, we used the results of our cDFT and NLPB calculations to determine ASAXS profiles as described in the \textit{Methods} section.
The results of these calculations for Rb$^+$ and Sr$^{2+}$, together with experimental data, are shown in Fig.~\ref{fig:rb-sr-asaxs}.
Similar results for CoHexCl$_3$ are shown in Fig.\ \ref{fig:cohex-asaxs}.

\section*{Discussion}

\subsection*{Comparison to Manning condensation}
The cDFT calculations of ionic distributions for the uniformly charged cylinder model (Fig.\ \ref{fig:cdft-cylinder}) reproduce the Manning condensation limits \cite{Manning:1978} with approximately 1M concentrations of singly-charged cations at the cylinder surface. A complete 3D cDFT model also reproduces the Manning condensation limit for monovalent cations: the {concentration of condensed Rb$^+$} on {the backbone}  and in minor groove is about 1 M (Fig.\ \ref{fig:cdft-Rb-helical}).
{The 1 M limit for monovalent ion concentrations at DNA surface corresponds to 76\% compensation of native B-form DNA charge by condensed counterions.
Manning's theory predicts that the concentration of condensed counterions is independent of bulk salt concentration in the range of 0.0001-–0.1 M and increases slightly for higher ionic strengths of monovalent electrolyte solution (80\% charge compensation for 0.5 M and 83\% for 1 M solutions).}
Additionally, the 3D cDFT model predicts that the multivalent ions form much denser layers at the DNA surface than the monovalent cations (Figures \ref{fig:cdft-Rb-helical}, \ref{fig:cdft-Sr-helical}, and \ref{fig:cdft-CoHex-helical}), consistent with Manning theory.
The good correlation between our non-mean-field cDFT model (with full ion-ion correlations included) and the mean-field Manning theory is somewhat surprising, particularly given the significant differences observed in the total condensed ion densities between cDFT and the mean-field NLPB approaches.
However, Manning theory indirectly accounts for interactions beyond first-order electrostatics through partitioning the total ion density into condensed ions and the surrounding ionic atmosphere.
This accounts for the success of Manning theory in predicting the {condensed concentrations} of 1:1 electrolyte counterions on DNA as observed in experiments \cite{Manning:1978} and recent MD simulations \cite{Giambasu:2014}.

\subsection*{Ion interaction with DNA grooves}
By definition, the cylinder model does not allow ion penetration inside DNA and therefore yields well-known monotonically decreasing counterion distributions shown in Fig.\ \ref{fig:cdft-cylinder}.
Thus, the model is not adequate for describing the interaction between DNA and small weakly solvated Na$^+$ and Rb$^+$ ions, which are known to penetrate into the minor grooves of DNA \cite{Giambasu:2014, Robbins:2014}.
However, our more detailed helical charge model allows ion penetration.
Simulations of RbCl solutions using this model showed that about half of the condensed Rb$^+$ ions are bound to the minor groove of the DNA molecule (Fig.\ \ref{fig:cdft-Rb-helical}).
The distributions of cation densities on the DNA {backbone}  and in the minor groove are highly structured: they exhibit a periodicity correlated with the periodic spacing of phosphate groups on {the DNA backbone} .
In contrast, cation distributions in the major groove are mostly featureless (see Fig.~S1 in Supporting Information), in agreement with previous simulations and experimental data \cite{Allahyarov:2003}.
Penetration of some cations into the grooves lowers the effective charge density on the DNA, limiting cation condensation on {the backbone} .

Increasing cation valency correlates with a stronger preference of cation binding to phosphate groups on the DNA {backbone}  (Figures \ref{fig:cdft-Rb-helical}, \ref{fig:cdft-Sr-helical}, and \ref{fig:cdft-CoHex-helical}).
A similar preference for CoHex$^{3+}$ binding to phosphates of B-DNA was also observed in MD simulations \cite{Tolokh:2014} and is determined by the strong electrostatic attraction of the trivalent cations to phosphate groups, CoHex$^{3+}$-CoHex$^{3+}$ repulsion, and steric inaccessibility of B-DNA minor groove to the large CoHex$^{3+}$ ions.
As shown in Figures \ref{fig:cdft-Sr-helical} and \ref{fig:cdft-CoHex-helical}, both Sr$^{2+}$ and CoHex$^{3+}$ ions preferentially bind to every fourth phosphate on the strand.
Further away from the DNA axis, the Sr$^{2+}$ density variations along the angular cylindrical coordinate have the same period as the period of the angular phosphate distribution {(Fig.~\ref{fig:cdft-Sr-helical})}.
The period of the density variations for CoHex$^{3+}$ {(Fig.~\ref{fig:cdft-CoHex-helical})} is two times larger than for Sr$^{2+}$; i.e., some Sr$^{2+}$ions can penetrate into the minor groove, while CoHex$^{3+}$ ions bind exclusively to phosphate groups on the {backbone} .

\subsection*{Influence of correlation on ion distributions}
To investigate the influence of ion correlation forces on the distribution of ions around DNA, we used a cDFT model without ion-correlation interactions (cDFT-nc) as well as the nonlinear Poisson-Boltzmann (NLPB) model, which also lacks correlation (Figures \ref{fig:cdft-Rb-helical}, \ref{fig:cdft-Sr-helical}, and \ref{fig:cdft-CoHex-helical}).
Both models without correlations yield qualitatively different ion distributions than the 3D cDFT calculations which include correlations.
In the presence of correlations, sterically allowed ions accumulate in the minor groove; in the absence of correlations, ions accumulate near phosphate groups on the exterior of the DNA strand.
The largest qualitative difference between NLPB and cDFT ion distributions was observed for the Rb$^+$ density distribution.
In NLPB, Rb$^+$ ions decorate the phosphate groups, driven by Coulombic interactions; the panoramic distribution of Rb$^+$ ions condensed on  {the backbone}  in NLPB model has a larger peak at 45$^\circ$ and a smaller one at 135$^\circ$ (Figures \ref{fig:cdft-Rb-helical}, \ref{fig:cdft-Sr-helical}, and \ref{fig:cdft-CoHex-helical}).
In contrast, ion-ion correlations reduce the effective electrostatic repulsion between cations promoting their penetration into the grooves.Due to stronger Coulomb interactions between multiply-charged cations the effect of correlations is weaker for Sr$^{2+}$ and CoHex$^{3+}$ resulting in the decrease in the fraction of counterions in the grooves with ion radius and charge (Fig.\ \ref{fig:cdft-Sr-helical} and Fig.\ \ref{fig:cdft-CoHex-helical}). For CoHex$^{3+}$ the concentration of counterions in the grooves becomes insignificant.As a result, NLPB and cDFT predict qualitatively similar panoramic density distributions on DNA {the DNA backbone}  for Sr$^{2+}$ and CoHex$^{3+}$ (Figures \ref{fig:cdft-Sr-helical} and \ref{fig:cdft-CoHex-helical}).

{As illustrated in the figures and Tables \ref{tab:models} and \ref{tab:results}, the} models without correlations (NLPB and cDFT-nc) are very similar to each other, indicating the major influence of correlation on even low charge-density (monovalent) ion behavior. This result contrasts the conclusion that correlations are insignificant in monovalent electrolytes from early theories of ion correlations  \cite{Grochowski.2008}. However, these theories considered electrolytes at uniformly charged surfaces ignoring the influence of the discreteness of charge distribution on fluctuations in ionic atmosphere. Not surprisingly, these models do not capture the experimentally observed attraction between like-charged polyelectrolytes in low concentration monovalent electrolytes \cite{Sedlak:1996, Manning:2011}. Recent molecular dynamic simulations also point to the importance of non-mean-field interactions between biomolecules and monovalent electrolytes manifested in a more structured ionic atmosphere than that predicted by NLPB \cite{Giambasu:2014, Robbins:2014}.
The small difference between the cDFT-nc and NLPB models (in the height of the double peak around 45$^\circ$) for Rb$^+$ is due to the solvent excluded-volume effects included in the cDFT-nc model and absent from NLPB theory (Fig.\ \ref{fig:cdft-Rb-helical}).

Ion correlations also influence ion-specific details in density distributions for counterions of the same valency.
3D cDFT results demonstrate that smaller Na$^+$ ions tend to accumulate on {the DNA backbone}  and minor groove while Rb$^+$ ions are more evenly distributed along the DNA helix (see Fig.\ \ref{fig:cdft-Rb-helical}a).
In the cDFT-nc and NLPB models with no correlation, the differences between Na$^+$ and Rb$^+$ distributions are significantly smaller (see Fig.\ \ref{fig:cdft-Rb-helical}b), suggesting that ion-correlation interactions are responsible for this effect.

\subsection*{Comparison with ASAXS experiments}
ASAXS profiles calculated using the 3D cDFT model show very good agreement with experimental data \cite{Andresen:2008} for RbCl solutions (Fig.\ \ref{fig:rb-sr-asaxs}).
The shapes of the scattering curves are very similar in the 1D cDFT and 3D cDFT models, with similar average numbers of condensed counterions:
the 1D and 3D cDFT calculations predict 34.9 and 34.6 condensed Rb$^{+}$ ions, respectively.
Both predictions are within error of the experimental measurement of 34 $\pm$ 3 ions \cite{Pabit:2010}.
However, the distribution of Rb$^+$ ions is different in these models:  all condensed cations decorate the cylinder surface (by definition) in the 1D model while half of the condensed cations are on {the DNA backbone}  and the other half are in minor grooves in the 3D models.
In contrast, the 3D NLPB model shows a significant deviation of the simulated scattering curve from the experimental one (Fig.\ \ref{fig:rb-sr-asaxs}).
As discussed in the previous section, penetration of some cations into DNA grooves reduces the negative electrostatic potential acting on cations in solution.
In the absence of any interactions beyond Coulomb forces, this penetration leads to lower concentrations of cations on the DNA surface and lower {concentrations of condensed counterions}.
Previous NLPB simulations demonstrated that adjusting the ionic radius of Rb$^+$ to its hydrated radius and prohibiting ion penetration into the DNA hydration shell can lead to closer agreement between 3D NLPB results and experiment \cite{Andresen:2004} -- but at the price of an incorrect ion distribution around DNA.

Comparison of calculated and experimental SrCl$_2$ data highlight the importance of {ion-solvent interactions} on ion distributions around DNA {(Fig.~\ref{fig:rb-sr-asaxs})}.
Sr$^{2+}$ ions have a significant hydration energy; approximately 3 times higher than that of monovalent alkali metal ions.
These strong cation-water interactions lower the entropy of water molecules around cations, but introduce a higher enthalpy cost for partial ion desolvation \cite{Lightstone:2001}.
Simulations with the solvent approximated as dielectric continuum do not account for such desolvation, limiting interactions in the system to first- and second-order electrostatic interactions: Coulomb and ion-correlation forces.
To understand the importance of these desolvation contributions, we used two variants of the cDFT model: one with attractive cation-water interactions and another without.
As shown in Fig.\ \ref{fig:rb-sr-asaxs}, ASAXS curves calculated using cDFT without desolvation contributions {via ion-solvent interactions} deviate significantly from the experimental data and the ASAXS curves calculated using the complete cDFT model.
On the other hand, the experimental ASAXS curves agree with those calculated from the complete cDFT model.
The importance of desolvation is also emphasized by the fact that inclusion of such interactions is essential for reproducing the chemical potentials of divalent cations but is not required for weakly hydrated alkali metal ions or CoHex$^{3+}$ (see Supporting Information).
In summary, {ion-solvent interactions} are important for accurately modeling ion-DNA interactions: desolvation reduces the excess chemical potential of cations and anions, lowering the effective concentration of electrolyte and weakening ion-ligand interactions.
Surprisingly, an NLPB model {that} includes neither {ion-solvent interactions} nor {ion-ion correlations} reproduces the experimental ASAXS curves for Sr$^{2+}$ (Fig.\ \ref{fig:rb-sr-asaxs}), although some differences are obvious in the more detailed radial {density} functions (Fig.\ \ref{fig:cdft-Sr-helical}).
This agreement is serendipitous and is due to cancellation of errors from the lack of ion-correlation, that favors ion accumulation in the grooves, and {ion solvation}, that limits ion concentration in the grooves.

Unexpectedly, the trivalent CoHexCl$_3$ solution is the simplest ion to model around DNA; CoHex$^{3+}$ can be reliably described by first-order electrostatics (i.e., direct Coulomb interactions).
CoHex$^{3+}$ ions decorate DNA {backbone} phosphates and do not penetrate inside B-DNA grooves {(Fig.~\ref{fig:cdft-CoHex-helical})}.
All models explored in this paper show reasonable agreement between the calculated ASAXS profiles and the experimental data (Fig.\ \ref{fig:cohex-asaxs}), and is consistent with all-atom MD simulations in explicit solvent \cite{Tolokh:2014}.
For these triply-charged ions, the good agreement between cDFT and NLPB is caused by the dominance of first-order electrostatics in ion-DNA interactions over higher-order ion-ion correlations.
Moreover, the large CoHex$^{3+}$ ionic diameter creates a steric barrier for ion penetration inside the grooves, rendering the 1D cylinder models adequate for calculating the average number of condensed CoHex$^{3+}$ ions.
Finally, because the diameter of CoHex$^{3+}$ is large, the field at its surface is comparable to Na$^+$ and the effects of {ion-solvent interactions} are lower than for the smaller divalent Sr$^{2+}$ ions. It follows from the current study that the models required to describe CoHex$^{3+}$ around a single B-DNA {backbone}  are relatively simple.
However, we expect that ion correlation forces will dominate DNA-DNA interactions between multiple {backbones}  due to the higher local phosphate charge density.

\section*{Conclusions}
We have studied the details of ionic atmospheres around DNA molecule for 1:1, 2:1 and 3:1 electrolytes using a combination of cDFT and NLPB methods.
Our calculations demonstrated that ion-ion correlation interactions induce counterion penetration into the DNA grooves, unless sterically prohibited by large ion radii.
In particular, ion binding in the grooves -- compared to binding on {the backbone}  -- has a profound effect on ion-induced nucleic acid condensation as demonstrated in our previous work \cite{Tolokh:2014}.
{Ion-solvent} interactions have an opposite effect:  when the enthalpy cost of desolvation is high (e.g., for Sr$^{2+}$ ions), ion-water interactions limit ion penetration into the DNA grooves.
Partial compensation of these two opposing effects explains the success of NLPB in reproducing the average number of condensed cations and the shape of the ASAXS  curves of the ion-counting experiments.
In contrast, cDFT model without ion-desolvation interactions was found to systematically overestimate ion concentration in DNA grooves.
Through the comparison of several cDFT models and experimental data, we demonstrated that a minimum model to describe ion-polyelectrolyte interactions should include long-range correlations arising from density and charge density fluctuations in electrolyte solution as well as short-range ion (de)solvation forces.
The latter interactions are often ignored in reduced models of electrolyte solutions limiting their applicability to the classes of weakly solvated ions.
Ion hydration forces are particularly pronounced in solutions of multiply-charged ions and give significant contribution to ion activity and, therefore, to ion-polyelectrolyte interactions.
Our results highlight important aspects of the properties of electrolyte solutions influencing ionic atmosphere around biomolecules that may significantly impact DNA condensation and biomolecules-ligand interactions.
{One caveat of the current work is its neglect of DNA sequence-specific effects which have been shown to influence ion binding in some cases \cite{Halle1998, Shui1998, McFail-Isom1999, Denisov2000, Pasi2015}.
The goal of our initial research was to understand the general characteristics of DNA-ion interactions that drive the behavior of different ionic species around DNA.
In the future, we plan to extend the DNA model to include such sequence-dependent structural variations.}

\section*{Author contributions}
MLS, DGT, and SAP performed the research, analyzed the data, and helped write the manuscript.  LP, AVO, and NAB designed the research, analyzed the data, and assisted in writing the manuscript.

\section*{Acknowledgments}
This work was supported by NIH Grant R01 GM099450.
NAB and MLS acknowledge fruitful discussions of hydration effects with Chris Mundy and Greg Schenter. The cDFT calculations were performed using PNNL Institutional Computing resources at Pacific Northwest National Laboratory. PNNL is a multiprogram national laboratory operated for DOE by Battelle under Contract DE-AC05-76RL01830.

\section*{SUPPLEMENTARY MATERIAL}

An on-line supplement to this article can be found by visiting BJ Online at \texttt{http://www.biophysj.org}.


\bibliographystyle{biophysj}
\bibliography{cdft}


\newpage

\listoffigures

\newpage

\listoftables



\end{document}